\documentclass[10pt,preprint,flushrt]{aastex}
\slugcomment{{\em Astrophys.\  J.\  Suppl.,} in press (February 2005)}

\shorttitle{Alfv\'{e}n Waves: Photosphere to Heliosphere}
\shortauthors{Cranmer and van Ballegooijen}

\begin{document}

\title{On the Generation, Propagation, and Reflection of
Alfv\'{e}n Waves from the Solar Photosphere to the Distant
Heliosphere}

\author{S. R. Cranmer and A. A. van Ballegooijen}
\affil{Harvard-Smithsonian Center for Astrophysics,
60 Garden Street, Cambridge, MA 02138}

\begin{abstract}
\baselineskip=11.7pt
We present a comprehensive model of the global properties of
Alfv\'{e}n waves in the solar atmosphere and the fast solar wind.
Linear non-WKB wave transport equations are solved from the
photosphere to a distance past the orbit of the Earth, and for
wave periods ranging from 3 seconds to 3 days.
We derive a radially varying power spectrum of kinetic and
magnetic energy fluctuations for waves propagating in both
directions along a superradially expanding magnetic flux tube.
This work differs from previous models in three major ways.
(1) In the chromosphere and low corona, the successive merging
of flux tubes on granular and supergranular scales is described
using a two-dimensional magnetostatic model of a network element.
Below a critical flux-tube merging height the waves are modeled
as thin-tube kink modes, and we assume that all of the kink-mode
wave energy is transformed into volume-filling Alfv\'{e}n waves
above the merging height.
(2) The frequency power spectrum of horizontal motions is
specified only at the photosphere, based on prior analyses
of G-band bright point kinematics.
Everywhere else in the model the amplitudes of outward and
inward propagating waves are computed with no free parameters.
We find that the wave amplitudes in the corona agree well
with off-limb nonthermal line-width constraints.
(3) Nonlinear turbulent damping is applied to the results of
the linear model using a phenomenological energy loss term.
A single choice for the normalization of the turbulent
outer-scale length produces both the right amount of damping
at large distances (to agree with in situ measurements) and
the right amount of heating in the extended corona (to agree
with empirically constrained solar wind acceleration models).
In the corona, the modeled heating rate differs by more than
an order of magnitude from a rate based on isotropic
Kolmogorov turbulence.
\end{abstract}


\section{Introduction}

Magnetic fields are known to play a significant role in
determining the equilibrium state of the plasma in the solar
atmosphere and solar wind (e.g., Parker 1975, 1991;
Narain and Ulmschneider, 1990, 1996; Priest 1999).
Much of the magnetic flux in the ``quiet'' photosphere seems to be
concentrated into small (100--200 km) intergranular flux tubes.
The physical processes that heat the chromosphere and corona
have not yet been identified definitively, but there is little
doubt that magnetohydrodynamic (MHD) effects are prevalent.
Even many proposed nonmagnetic mechanisms depend on the
underlying properties of the magnetically structured atmosphere.
The outflowing solar wind is fed by open magnetic flux tubes,
and many MHD processes have been proposed to deposit heat and
momentum at locations ranging from the extended corona to
interplanetary space.

The continually evolving convection below the photosphere
gives rise to a wide spectrum of MHD fluctuations in the
magnetic atmosphere and wind.
The propagation of waves through the solar atmosphere has been
studied for more than a half century (Alfv\'{e}n 1947;
Schwarzschild 1948; Biermann 1948).
Although compressible (e.g., acoustic and magnetoacoustic) MHD
waves are likely to be dynamically and energetically important
in some regions of the atmosphere, it is the mainly
incompressible Alfv\'{e}n mode that has been believed for many
years to be dominant in regions that are {\em open to the
heliosphere} (e.g., Osterbrock 1961;
Kuperus, Ionson, \& Spicer 1981).
Indeed, the MHD fluctuations measured by spacecraft in the
solar wind have a strongly Alfv\'{e}nic character
(Belcher \& Davis 1971; Hollweg 1975; Tu \& Marsch 1995;
Goldstein, Roberts, \& Matthaeus 1995).

Even though much has been learned about the generation,
propagation, reflection, and damping of Alfv\'{e}n waves
in the solar atmosphere, most earlier studies have focused
on only a finite range of heights and treated the interactions
with regions above or below as boundary conditions.
This necessarily involved the approximation that the relevant
phenomena are mainly local, i.e., that they do not depend
on the conditions very far away from the region being modeled.
There are circumstances, however, for which this approximation
breaks down.
For example, the properties of reflecting Alfv\'{e}n waves with
long periods (i.e., of order 1 day) in the solar wind depend
formally on the conditions ``at infinity,'' since they behave
asymptotically as standing waves (e.g., Heinemann \& Olbert 1980).

In this paper we construct a comprehensive model of the
radially evolving properties of Alfv\'{e}nic fluctuations
in a representative open magnetic region of the solar
atmosphere and fast solar wind.
The model takes account of nonlocal interactions by tracing
the waves from their origin as transverse flux-tube oscillations
in the photosphere all the way out to the interplanetary
medium (truncated for convenience at 4 AU).
This is done with the smallest possible number of free parameters.
There are two overall aims of this paper:
\begin{enumerate}
\item
We wish to understand better the global ``energy budget'' of
Alfv\'{e}n waves, including relative amplitudes of inward and
outward propagating waves, along the open flux tubes that feed
the solar wind.
\item
In order to determine how MHD turbulence contributes to
the heating of the extended solar corona, we need to pin down
precisely how Alfv\'{e}n waves provide the natural
preconditions for a turbulent cascade.
\end{enumerate}
The second aim above was motivated by a recent study of the
small-scale dissipation of MHD turbulence in the extended
corona (Cranmer \& van Ballegooijen 2003).
This kinetic dissipation depends strongly on how the turbulence
is excited at its largest scales, and in this paper we attempt
to put the ``outer scale'' wave dynamics on firmer footing so
that the resulting ``inner scale'' can be better understood.
The work described by this paper builds on prior studies by
Hollweg  (1973, 1978a, 1981, 1990), Heinemann \& Olbert (1980),
Spruit (1981, 1984), An et al.\  (1990), Barkhudarov (1991),
Velli (1993), Lou (1993, 1994), Lou \& Rosner (1994),
MacGregor \& Charbonneau (1994), Kudoh \& Shibata (1999),
Matthaeus et al.\  (1999), Hasan et al.\  (2003),
and many others.

One unique aspect of this paper is that the photospheric spectrum
of transverse fluctuations---which constrains the Alfv\'{e}n wave
amplitudes at all larger radii---is specified directly from
detailed observations of magnetic bright point (MBP) motions and
is not (as in many other models) given as an arbitrary boundary
amplitude.
However, a complete physical description of the fluctuations in
the photosphere (e.g., the relative inward/outward amplitudes
and the kinetic/magnetic energy partition) is obtained only
after the fully nonlocal wave reflection has been computed for
all radii.
Another way this work differs from many previous models is
that the expansion and successive merging of flux tubes on
granular and supergranular scales is described using a
two-dimensional model of a magnetic network element in the
inhomogeneous solar atmosphere.

Despite the attempted comprehensiveness of this model, we
needed to make three specific approximations in order to
render the calculations tractable.
These approximations are summarized here, but they are also
discussed further below and justified for certain regimes of
applicability.
First, we ignore all effects of compressible fluctuations (e.g.,
acoustic waves; fast and slow magnetosonic waves) despite
their importance in understanding observed intensity
oscillations and chromospheric heating.
This is done in order to fairly assess the relative importance
of the incompressible Alfv\'{e}n mode before resorting to
more involved models.
Second, the Alfv\'{e}n wave model is mainly linear,
which limits its applicability in regions where the wave
amplitudes become large in comparison to background field
strengths and characteristic speeds.  (Some effects
of nonlinearity are examined, though, in {\S}~6.)
Third, we do not model explicitly the back-reaction of the
waves on the mean properties of the solar atmosphere and
solar wind.
We do, however, compute quantities such as the
wave pressure acceleration and turbulent heating rate
for use in future models of this back-reaction.

The remainder of this paper is organized as follows.
In {\S}~2 we give an overview of the physical processes
to be incorporated in the wave models together with a
``cartoon'' description of the steady-state magnetic field
topology.
In {\S}~3 we describe the adopted steady-state (i.e.,
zero-order) plasma conditions in detail.
The specification of the photospheric frequency spectrum
of transverse fluctuations is given in {\S}~4, and
the wave equations to be solved are given in {\S}~5.
Solutions of these equations, including some with
various prescriptions for damping, are presented
in {\S}~6.
We conclude with a summary of major results ({\S}~7)
and a discussion of remaining issues ({\S}~8).
Appendix A contains supplementary equations describing
analytic solutions of the kink-mode wave equations for
isolated flux tubes in an isothermal atmosphere.
Appendix B compares various published formalisms for
the non-WKB transport equations for Alfv\'{e}n waves in an
accelerating wind.
Appendix C summarizes the properties of the fully developed
anisotropic MHD turbulence spectrum discussed by, e.g.,
Cranmer \& van Ballegooijen (2003).

\section{Overall Picture of Open Field Regions}

Figure 1 illustrates the basic magnetic field geometry that
we believe is representative of flux tubes that feed the
high-speed solar wind.
A key feature of the adopted configuration is the successive
merging of strong-field magnetic flux tubes between granules
(Figure 1a) and supergranules (Figure 1b).
On the largest scales, Figure 1c shows the more or less
axisymmetric magnetic field that is characteristic of solar
minimum (using the field model of Banaszkiewicz, Axford,
\& McKenzie 1998), but nearly all of the work presented in
this paper can also be applied straightforwardly to open-field
regions at other phases of the solar cycle.

We assert that most of the plasma that eventually becomes the
time-steady solar wind originates in intergranular magnetic flux
tubes known observationally as G-band bright points,
network bright points, or in groups as ``solar filigree.''
This assertion is seemingly uncontroversial (knowing what
we do about solar magnetic fields), though it is seldom stated.
Adopting a ripening convention in nomenclature, we
refer to these 100--200 km size photospheric features as
{\em magnetic bright points} (MBPs).\footnote{We
also note that MBPs are not the same phenomena as the
larger K$_{2V}$ bright points in the chromosphere
(Rutten \& Uitenbroek 1991) or the still larger
X-ray bright points in the low corona
(e.g., Golub et al.\  1977; Parnell 2002).}
High-resolution observations reveal the presence of MBPs
in the dark lanes between granules,
and these features are associated with regions of strong
(1--2 kG) magnetic field believed to be contained within nearly
vertical flux tubes (e.g., Sheeley 1967; Dunn \& Zirker 1973;
Muller 1983, 1985; Piddington 1978; Rabin 1992; Solanki 1993;
Berger \& Title 2001).
There is increasing evidence for magnetic structures on even
smaller scales than 100--200 km, but we leave the study of these
structures to future work (see, e.g., Stein \& Nordlund 2002;
S\'{a}nchez Almeida, Emonet, \& Cattaneo 2003;
Trujillo Bueno, Shchukina, \& Asensio Ramos 2004).

As magnetic flux rises stochastically from the convection zone
to the photosphere (e.g., Priest, Heyvaerts, \& Title 2002), 
field lines are simultaneously jostled horizontally by fluid
motions on granular (1--2 Mm) scales.
Magnetic flux is concentrated into thin tubes by some combination
of ``flux expulsion'' from the upflowing granule centers to
the downflowing lanes (Parker 1963), the rapid evacuation
of these downflowing superadiabatic regions (i.e., convective
collapse; Parker 1978; Spruit 1979), and enhanced radiative
cooling leading to thermal relaxation (S\'{a}nchez Almeida 2001).
MBPs are observed frequently to merge with neighboring flux
elements and to spontaneously fragment into several pieces
(e.g., Berger et al.\  1998).
Once formed, MBPs continue to be shaken back and forth by the
underlying convective motions (Kulsrud 1955; Osterbrock 1961;
van Ballegooijen 1986; Huang, Musielak, \& Ulmschneider 1995),
which results in various kinds of wavelike fluctuations
describable in the ``thin-tube'' MHD limit (Spruit 1981,
1982, 1984; see further references in {\S}~5.1 below).
Oscillatory motions can also be induced by impulsive
reconnection events (e.g., Moore et al.\  1991) or, conversely,
the random wave trains may display observational time signatures
that could be misinterpreted as small-scale flaring
(Moriyasu et al.\  2004).

There is a great variety of MHD wave activity expected and
observed in the inhomogeneous solar atmosphere.
In addition to isolated MBP fluctuations, there is ambient
acoustic wave energy excited by convection,
shock steepening in the chromosphere and transition region,
and both driven and free oscillations in sunspots and coronal
loops (see recent reviews by Axford et al.\  1999;
Roberts 2000; Bogdan 2000; Hirzberger 2003).
A significant fraction of the Sun's magnetic flux may also
be distributed {\em outside} the MBPs, network, and active
regions (e.g., Schrijver \& Title 2003), and acoustic waves
in ``field-free'' regions may in fact be magnetoacoustic.
Damping of MHD waves and turbulence has been a key ingredient
in many proposed models of chromospheric and coronal heating.
Our main focus, though, is on the incompressible waves that
eventually escape from the atmosphere into the solar wind.

Magnetic flux tubes rooted in MBPs undergo both transverse
(kink-mode) and longitudinal (sausage-mode) oscillations that
can propagate upward from the photosphere.
Because the strong-field flux tubes are in horizontal pressure
equilibrium with the surrounding weak-field material, they
have a lower density and thus are susceptible to buoyancy
effects and evanescence for long enough periods.
Like acoustic waves, the compressible longitudinal modes
steepen into shocks and damp over a few scale heights, while
the incompressible kink modes can propagate into the corona 
relatively undamped.
Nonlinear effects can lead to mode conversion between kink
and longitudinal modes (e.g., Ulmschneider, Z\"{a}hringer,
\& Musielak 1991).

Somewhere in the low chromosphere, the thin flux tubes are
believed to expand laterally to the point where they merge
with one another into a homogeneous network field distribution
(Spruit 1984; Pneuman, Solanki, \& Stenflo 1986;
Gu et al.\  1997).
At this point, the thin-tube description breaks down and
standard MHD wave theory becomes more applicable
(i.e., kink modes become transverse Alfv\'{e}n waves).
The merged network flux bundles have horizontal scale lengths
of 2 to 6 Mm and are probably maintained by large-scale
convective flows that push the field to the edges of
supergranular cells.
At a larger height---still below the chromosphere-corona
transition region---the network magnetic field expands
laterally to 10--30 Mm scales and is thought
to merge again into a large-scale ``canopy''
(e.g., Kopp \& Kuperus 1968; Gabriel 1976; Giovanelli 1980;
Anzer \& Galloway 1983; Dowdy, Rabin, \& Moore 1986).
The spatial scale of the canopy is set by the typical distance
between network flux bundles (i.e., the size of supergranulation
cells) in the chromosphere.
Observational evidence for preferential wind acceleration
in the rapidly expanding network ``funnels'' is growing
(Rottman, Orrall, \& Klimchuk 1982; Hassler et al.\  1999;
Peter \& Judge 1999; Aiouaz et al.\  2004;
see also Mart\'{\i}nez-Galarce et al.\  2003)
but is still not definitive (e.g., Dupree, Penn, \& Jones 1996). 

As waves propagate upward into the corona, the radially
varying Alfv\'{e}n speed allows for gradual linear reflection
(Ferraro \& Plumpton 1958).
The transition region can also act a sharp ``reflection barrier''
to Alfv\'{e}n waves with wavelengths exceeding the local scale
length of the Alfv\'{e}n speed in that thin zone (see {\S}~6).
It is thus possible for a time-steady superposition of upward
and downward propagating Alfv\'{e}n waves to be maintained
(e.g., Hollweg 1981, 1984).
Strong reflection is not necessarily an impediment to there
being a substantial upward wave flux; one merely needs more
power in the upward modes than in the downward modes.
Somewhere in the solar atmosphere the MHD fluctuations become
turbulent, but it is unclear whether the turbulent cascade
becomes energetically important in the photosphere
(Petrovay 2001), in the chromosphere and transition region
(Chae, Sch\"{u}hle, \& Lemaire 1998),
or in the corona (e.g., Dmitruk \& Matthaeus 2003).

In the extended corona, the high-speed solar wind begins to
accelerate supersonically above a heliocentric distance of
2 to 3 solar radii ($R_{\odot}$).
In the radially inhomogeneous wind, the dissipationless
propagation of MHD waves does work on the mean fluid and
provides an added wave-pressure acceleration (e.g.,
Belcher 1971; Jacques 1977; Leer, Holzer, \& {Fl\aa} 1982).
The presence of the wind also modifies how waves propagate,
and above the Alfv\'{e}n critical point---where the wind speed
equals the local Alfv\'{e}n speed at about 10 $R_{\odot}$---both
the inward and outward modes are advected outward with the wind.
Large coronal holes are the most probable source regions
for the fast solar wind (for a review of observations and
theoretical models, see Cranmer 2002).
Some flux tubes in coronal holes have a higher density than
the surrounding open-field plasma; these ``polar plumes''
seem to trace out the superradial expansion of the
merged-canopy magnetic field in the corona (e.g.,
DeForest, Lamy, \& Llebaria 2001).

In addition to Alfv\'{e}n waves, there is some evidence for
both fast-mode and slow-mode magnetosonic waves in the corona
(Ofman, Nakariakov, \& DeForest 1999; Nakariakov et al.\  2004),
but they have been observed only in relatively confined regions
such as loops and plumes.
Fast and slow modes are believed to be more attenuated by
collisional damping processes than Alfv\'{e}n waves before they
reach the corona.
However, once some fraction of the energy flux of Alfv\'{e}n
(and possibly fast-mode) waves escapes into the solar wind,
the classical transport theory of collisional damping begins
to break down and collisionless wave-particle interactions
should dominate the damping (e.g., Hollweg \& Isenberg 2002;
Marsch, Axford, \& McKenzie 2003; Cranmer 2000, 2001, 2002, 2004;
and many references therein).
Studies of wave-particle resonance damping have seen a recent
resurgence of interest because of their potential importance in
producing the preferential ion heating and acceleration seen both
in the extended corona (Kohl et al.\  1997, 1998, 1999)
and in~situ (Marsch 1999).

As Alfv\'{e}n waves propagate into interplanetary space, their
velocity and magnetic-field amplitudes grow to nonlinear
magnitudes (i.e., $\delta B / B_{0}$ becomes of order unity).
Higher order ponderomotive effects begin to dominate the wave
propagation equations (Lau \& Siregar 1996) and collisionless
wave-wave resonances can create additional mode coupling and
damping (Lee \& V\"{o}lk 1973; Lacombe \& Mangeney 1980;
Lou 1993).
At some point it may even be inappropriate to use a
``wavelike'' paradigm to discuss the increasingly turbulent
fluctuations (e.g., Goldstein et al.\  1995).
In the outer heliosphere, the description of magnetic flux
tubes as relatively closed systems---implicit in the above
discussion---breaks down as well, since processes such as
stream-stream interactions, interstellar pickup ion injection,
and cosmic ray transport increasingly dominate the physics.

In this paper we purposefully examine only a subset of the
many processes listed above.
This is done mainly to keep the modeling tractable, but
it also helps clarify the extent to which the {\em specific
wave modes that we study} can account for various observations.

\section{Steady-State Plasma Conditions}

Models of linear waves depend sensitively on the assumed
background zero-order plasma properties (e.g., density,
flow speed, magnetic field strength, flux tube geometry).
In this section we describe the empirically constrained
time-steady plasma along the central axis of a radially
pointed (but superradially expanding) magnetic flux tube,
from the photosphere to a distance of 4 AU.
The empirical description consists of two parts:
a two-dimensional magnetostatic model of thin MBP flux tubes
that expand into the supergranular network canopy ({\S}~3.1)
and a one-dimensional analytic continuation of the plasma
parameters in the extended corona, assumed here to be along
the axis of symmetry of a polar coronal hole at solar
minimum ({\S}~3.2).

\subsection{Network Magnetic Structure}

We first develop a two-dimensional numerical model of a
supergranular network element as a collection of thin flux tubes.
The gas pressure in the atmosphere decreases with increasing
height, causing a lateral expansion of the flux tubes.
Neighboring MBP flux tubes within the network element merge
into a monolithic structure at some height $z_m$.
(Heights $z$ are measured from the optical-depth-unity
photosphere; radii $r$ are measured from Sun center.)
For a thin, isothermal flux tube in pressure equilibrium
with its surroundings, the interior magnetic field strength
varies with height as $e^{-z/2H}$, where $H$ is the pressure
scale height (e.g., Spruit 1981, 1984).
Thus we determine the so-called {\em merging height}
by solving for
\begin{equation}
  z_{m} \, \approx \, 2H \ln ( B_{\odot} / \bar{B} )
  \,\,\, ,
\end{equation}
where $B_{\odot}$ is the field strength inside the flux tube
at the base of the photosphere ($z=0$) and $\bar{B}$ is the
average flux density in the network patch.
Observations of the MBP flux-tube field strength range between
1000 and 2000 G, and the network-averaged field strength
varies between about 20 and 300 G (e.g., Gabriel 1976;
Giovanelli 1980).
In order to determine a representative value for the merging
height we use $H = 120$ km, $B_{\odot} = 1500$ G, and
$\bar{B} = 100$ G to obtain $z_{m} \approx 600$ km, a height
in the low chromosphere (see also Pneuman et al.\  1986).
Above the merging height the network element consists of a single
thick flux tube that further expands with height.
The outer edge of this tube forms a magnetic canopy that
overlies the neighboring supergranular cells.
A second merging occurs when neighboring network elements come
together at a ``canopy height'' $z_c$ above the supergranular cell
centers; we set $z_{c} = 1$ Mm (see also Hasan et al.\  2003).
Figure 2 shows the three-part structure of the network element:
\begin{enumerate}
\item
The region below the merging height ($0 < z < 0.6$ Mm) is
described as a collection of thin flux tubes embedded in a
field-free medium.
The field strength $B(z)$ is assumed to be the same for all
flux tubes, and their cross sections and other plasma properties
are consistent with the thin-tube approximation.
\item
Between heights of 0.6 and 1 Mm, the merged network flux element
expands laterally to the edges of the supergranular cell, which
overlies a field-free cell-center chromosphere.
\item
Between 1 and 12 Mm, the ``fully merged'' magnetic field fills
the supergranular cell volume and expands primarily in the
vertical direction.
\end{enumerate}
In the remainder of this paper, the term ``merging height''
refers specifically to the merging of the thin flux tubes at
$z_{m} = 0.6$~Mm.

The total magnetic flux $\Phi$ of the network element is
constrained empirically to be $3.7 \times 10^{19}$ Mx.
The MBP field strengths at the photosphere and at the merging
height are 1430 G and 120.4 G, respectively.\footnote{%
The magnetic field is assumed to point outward---i.e., it is
assumed to be of northern polarity---but the physics is the same
for an inward pointing field.}
The upper, monolithic part of the magnetic structure ($z > z_m$)
is assumed to be cylindrically symmetric and is described by a
modified version of the magnetostatic flux tube model of
Hasan et al.\  (2003).
In the present case, the flux tube is contained within a cylinder
of transverse radius $s_{0} = 10$ Mm, which simulates the effect
of neighboring network elements.
The internal and external gas pressures are taken from
semiempirical VAL/FAL models by Vernazza, Avrett, \& Loeser (1981)
and Fontenla, Avrett, \& Loeser (1990, 1991, 1993, 2002).
The magnetic field components $B_{s}(s,z)$ and $B_{z}(s,z)$ in
cylindrical geometry (using $s$ as the perpendicular distance
from the central flux tube axis) are computed by varying the
shapes of the field lines until a minimum-energy state is
obtained (for details see Hasan et al.\  2003).
The horizontal distribution of $B_{z}(s,z_{m})$ at the merging
height is adjusted iteratively such that the magnitude of the
field $B(s,z_{m})$ is independent of $s$.
This is needed for consistency with the constant field strength
of the thin flux tubes just below the merging height.

Figure 2 shows the final iterated magnetic structure.
The cross sections of the flux tubes below the merging height
($z < 0.6$ Mm) were computed by matching $B_s$ and $B_z$ at the
merging height and assuming that $B_s$ is independent of height
within each flux tube (different tubes have different $B_s$).
The transverse radius of the ``network patch'' in the photosphere
is about 3 Mm.
We can estimate the number $N$ of MBPs inside the approximately
circular patch, but we note that the wave analysis below does
not depend on the value of $N$.
From the conservation of magnetic flux, we know that the
filling factor of MBP flux tubes at the photosphere
is given by $B(z_{m})/B_{\odot} \approx 0.08$.
Observationally, the transverse radii of MBPs at the photosphere
are 50 to 100 km.
The number of flux tubes required to fill an area
that is 8\% of the total patch area ($A = \pi s_{\odot}^2$,
with $s_{\odot} = 3$ Mm) ranges between $N \approx 75$
(for the upper limit on MBP radius) and $N \approx 300$
(for the lower limit).
We can also define a mean distance between nearest neighbors
$d_{nn}$ by dividing up up the network patch into $N$
equal-area circles and defining $d_{nn}$ as twice their radii.
Thus, $d_{nn} = 2 s_{\odot} N^{-1/2}$, which gives values
of 350 and 700 km for the above limiting cases.

Before continuing, it is worthwhile to ask the question:
``Why is it not possible to extend the `monolithic' flux-tube
model all the way down to the photosphere?''
Observationally, the photosphere is far from monolithic in
its distribution of magnetic fields.
Granulation causes both the formation of thin MBP flux tubes
and their spreading out (via random walk).
The larger-scale supergranulation pushes the magnetic
elements back together into network lanes and vertices.
This competition between MBP spreading and convergence leads
to a {\em dynamical equilibrium} described by the different
assumptions applied above and below the merging height.
A more complete model must contain physics that naturally
captures this equilibrium state, but hopefully without the
abrupt transition assumed here at $z_m$.

\subsection{Superradial Expansion in the Solar Wind}

Here we describe the plasma parameters along the central axis
of the modeled network element at heights ranging from 12 Mm
(the top of the magnetostatic model) to interplanetary space.
We adopt a slightly modified version of the polar magnetic
field configuration of Banaszkiewicz et al.\  (1998):
\begin{equation}
  \frac{B_0}{1.789 \mbox{\, G}} \, = \,
  \frac{2}{x^{3}} + \frac{4.281}{x^{5}} +
  \frac{1}{a_{c} (x + a_{c})^2} +
  \frac{0.67}{\exp [ 384 (x - x_{0} ) ]}
\end{equation}
where $B_0$ is now defined as the radial component of the
field along the axis of symmetry, $x = r / R_{\odot}$, and
the above description applies only for $x \geq x_{0}$, with
$x_{0} \equiv 1.0172$ (i.e., the top height [12 Mm] of the
magnetostatic grid).
The model above uses the same current-sheet constant
($a_{c} = 1.538$) used by Banaszkiewicz et al.\  (1998), but
with a 5\% modification to their preferred quadrupole constant.
We also add an exponential correction term (which drops rapidly
to nearly zero for $x \gtrsim 1.03$) to ensure that the value
and slope of $B_0$ match those of the magnetostatic model
at $x_0$.
Figure 3a plots the product $x^{2} B_0$ versus height both
below and above $x_0$.
This quantity is proportional to the inverse of the
traditional superradial divergence factor $f$, and it
shows the outer monopolar expansion region
($B_{0} \propto r^{-2}$) as a constant.
For comparison we also plot the analytic functions used in
the funnel models of Hackenberg, Marsch, \& Mann (2000) and
Li (2003).

For the radial dependence of the electron density, we use
a function motivated by fits to white-light polarization
brightness measurements in the extended corona:
\begin{equation}
  \frac{n_e}{1.3 \times 10^{5} \mbox{\, cm}^{-3}} \, = \,
  \frac{1}{x^{2}} + \frac{25}{x^{4}} + \frac{300}{x^{8}} +
  \frac{1500}{x^{16}} + \frac{5796}{x^{33.9}}
\end{equation}
which applies only for $x \geq x_{0}$.
The overall scale set by the inverse square term was
adjusted to match in~situ density measurements at 1 AU
(i.e., $x = 215$).
The middle three terms above were adjusted to produce
agreement with measurements by, e.g.,
Guhathakurta \& Holzer (1994), Fisher \& Guhathakurta (1995),
Doyle, Teriaca, \& Banerjee (1999), Esser \& Sasselov (1999),
and Figure 10 of Lie-Svendsen, Hansteen, \& Leer (2003).
The last term was set to match the value and slope of
the magnetostatic model density at $x_0$.
Figure 3b plots the hydrogen number density $n_{\rm H}$,
computed assuming a helium-to-hydrogen number density ratio
of 0.05 (i.e., $n_{e} = 1.1 \, n_{\rm H}$, with the total
mass density $\rho$ given by $1.2 \, n_{\rm H} m_{\rm H}$).

Figure 3c shows several velocity quantities along the
central axis of the flux tube.
The hydrogen outflow speed $u$ was computed by mass flux
conservation (i.e., $\rho u / B_0 =$~constant), with the
mass loss rate set by {\em Ulysses} measurements in
interplanetary space (Goldstein et al.\  1996).
At 1 AU, the product $u n_{\rm H} = 2 \times 10^8$ cm$^{-2}$
s$^{-1}$, and thus we compute $u = 781.2$ km s$^{-1}$ at 1 AU.
At an infinite distance, $u$ approaches a constant value of
781.9 km s$^{-1}$.
The Alfv\'{e}n speed, defined as
\begin{equation}
  V_{A} \, \equiv \, B_{0} / \sqrt{4\pi\rho}
  \,\,\, ,
\end{equation}
is nearly constant below the merging height (and would
be precisely constant for an isothermal atmosphere)
and rises to two successive maxima:  2530 km s$^{-1}$
at $r = 1.004 \, R_{\odot}$, and 2890 km s$^{-1}$
at $r = 1.53 \, R_{\odot}$.
The Alfv\'{e}n speed drops to 31.3 km s$^{-1}$
at 1 AU and decreases nearly exactly as $1/r$ after that.
The Alfv\'{e}n critical point, where $u = V_A$,
is at $r_{A} = 9.70 \, R_{\odot}$.

\section{Photospheric Fluctuation Spectrum}

The lower boundary condition for our model of Alfv\'{e}nic
fluctuations is the power spectrum of transverse MBP motions
in the photosphere.
The dynamical behavior of G-band bright points has been
studied observationally by a number of groups
(see, e.g., Muller et al.\  1994; Berger \& Title 1996;
van Ballegooijen et al.\  1998; Berger et al.\  1998;
Krishnakumar \& Venkatakrishnan 1999; Nisenson et al.\  2003).
In this section we present an empirical description of
MBP dynamics as a linear superposition of two types of
motion: (1) the ``random walk'' undertaken by isolated
flux tubes, and (2) a series of rapid ``jumps'' that
occur when individual flux tubes merge, fragment, or
reconnect with surrounding magnetic field.

The general procedure for specifying the power spectrum of
horizontal MBP kinetic energy (as a function of frequency)
is illustrated in Figure 4.
The primary measured quantity is a time series of discrete
position measurements for the MBPs which can be
differentiated to obtain horizontal velocity components
$v_{x}(t)$ and $v_{y}(t)$, with the $z$ direction being normal
to the solar surface.
(Observations support the rational assumption that there is
no preferred global direction on granular scales, so below we
discuss just the $v_x$ component and assume that $v_y$ is
statistically equivalent.)
The timescale dependence of MBP fluctuations is encapsulated
in the velocity autocorrelation function, which we define
for a time sequence of $N$ measured velocities $v_{x,n}$ as
\begin{equation}
  C_{xx,m} \, = \, \frac{1}{N-m} \sum_{n=1}^{N-m}
  v_{x,n} \, v_{x,n+m}
\end{equation}
for the time index $n$ between 1 and $N$ and an arbitrary
delay time represented by index $m$.
This expression is the discrete version of the more general
definition
\begin{equation}
  C_{xx} (\tau) \, = \, \lim_{\Delta t \rightarrow \infty}
 \frac{1}{\Delta t} \int_{-\Delta t/2}^{+\Delta t/2} dt \,
  v_{x}(t) \, v_{x}(t+\tau)
  \label{eq:autoint}
\end{equation}
for delay time $\tau$.
The unidirectional power spectrum $P_x$ (i.e., $|v_{x}|^{2}$
per unit interval of frequency $\omega$) is the Fourier
transform of the autocorrelation function, with
\begin{equation}
  P_{x} (\omega) \, \equiv \, \frac{1}{2\pi}
  \int_{-\infty}^{+\infty} d\tau \, C_{xx}(\tau) \,
  e^{i \omega \tau}
\end{equation}
(e.g., the Wiener-Khinchin theorem; see also
van Ballegooijen et al.\  1998).
Note that $P_{x}$ is defined for both positive and negative
$\omega$.
For the present applications, all functions are symmetric
about zero frequency, and we will thus consider only positive
frequencies (simultaneously doubling the normalization of
$P_x$ to conserve total energy).
In general, we specify the kinetic energy power spectrum
$P_K$, which is defined in such a way as to integrate to
the total kinetic energy density $U_K$ in transverse motions:
\begin{equation}
  U_{K} \, \equiv \, \frac{\rho \langle \delta V \rangle^2}{2}
  \, = \,
  \int_{0}^{\infty} d\omega \, P_{K} (\omega)
  \,\,\, ,
  \label{eq:UKearly}
\end{equation}
with
\begin{equation}
  P_{K} (\omega) \, \equiv \, \frac{\rho}{2}
  \left[ 2P_{x}(\omega) + 2P_{y}(\omega) \right]
  \, = \, 2\rho P_{x} (\omega)  \,\,\, .
  \label{eq:Pxy}
\end{equation}
The factors of two inside the square brackets take account
of the negative frequencies, and
the last expression assumes $P_{x} = P_{y}$.

In the analysis below we derive power spectrum components
for the two assumed phases of the MBP motion: the random
walk (subscript $w$) of isolated flux tubes, and the occasional
discrete jumps (subscript $j$) caused by merging, fragmenting,
or reconnecting.
Specifying the power separately for these two phases is not
the ideal solution, but it is all that can be done at present.
Ideally, the proper observational procedure would be to
determine the {\em complete} time series $v_{x}(t)$ for
the walk and jump phases taken together, then compute the
autocorrelation function and total power spectrum
consistently.
Unfortunately, because MBPs fragment or merge during
the jump phases, it is extremely difficult to ``follow''
a single feature during these times to determine the
complete time series.

The random-walk component of MBPs, during the times they exist
as separate entities, was studied by van Ballegooijen
et al.\  (1998) and Nisenson et al.\  (2003).
These observations yielded the result that the discretely derived
autocorrelation functions can be fit well by Lorentzian functions
(see Figure 4e), with
\begin{equation}
  C_{xx,w} (\tau) \, = \, \frac{\sigma_{w}^2}
  {1 + (\tau / \tau_{w})^2}
\end{equation}
and the more precise observations of Nisenson et al.\  (2003)
gave values of $\sigma_{w}^{2} \approx 0.8$ km$^{2}$ s$^{-2}$
and $\tau_{w} \approx 60$ s.
We adopt these values in the models below.
Thus, the walk-component of the kinetic energy spectrum is
given by
\begin{equation}
  P_{K,w}(\omega) \, = \, \rho \sigma_{w}^{2} \tau_{w}
  e^{-\tau_{w} \omega} \,\,\, .
\end{equation}
This is plotted in Figure 4g, but note that all power spectra
are plotted as the product $\omega P (\omega)$ because this
denotes the power per decade of frequency (i.e., the energy
density per unit $\log \omega$).
Maxima in this quantity highlight the frequencies that
contribute most to the total wave energy.

The impulsive ``jump'' phase of MBP motions is described by
Choudhuri, Dikpati, \& Banerjee (1993), Berger et al.\  (1998),
Hasan, Kalkofen, \& van Ballegooijen (2000), and others.
The potential for rapid transitions in the locations of
thin flux tubes is also indicated in empirical models of the
quasi-equilibrium evolution of granular magnetic fields
(van Ballegooijen \& Hasan 2003), in which slow motions of
separatrix surfaces in the photosphere are amplified at
larger heights due to flux-tube expansion.
We model an impulsive MBP event as narrow Gaussian
enhancement in $v_{x}(t)$ centered on an arbitrary $t=0$.
A series of these events is assumed to occur with a mean time
interval $\Delta t$ between events.\footnote{%
Although we believe a constant interval $\Delta t$ captures
the essential nature of these jumps and their contribution to
the energy spectrum, a more accurate way of modeling them
would be to sample from an empirically derived ``waiting-time''
probability distribution.}
The Gaussian jump, with velocity amplitude $\sigma_j$ and
$1/e$ half-life $\tau_j$, is described by
\begin{equation}
  v_{x} (t) \, = \, \sigma_{j} e^{-(t / \tau_{j})^{2}}
\end{equation}
between times $-\Delta t / 2$ and $+\Delta t / 2$, and the
limit $\Delta t \rightarrow \infty$ in eq.~(\ref{eq:autoint})
is not taken.
The autocorrelation function is thus
\begin{equation}
  C_{xx,j} (\tau) \, = \, \sqrt{\frac{\pi}{2}} \,
  \frac{\sigma_{j}^{2} \tau_{j}}{\Delta t} \exp
  \left( - \frac{\tau^2}{2 \tau_{j}^2} \right)
\end{equation}
(see Figure 4f) and the resulting kinetic energy spectrum is
given by
\begin{equation}
  P_{K,j}(\omega) \, = \, \frac{\rho \sigma_{j}^{2} \tau_{j}^{2}}
  {\Delta t} \exp \left( - \frac{\omega^{2} \tau_{j}^{2}}{2}
  \right)
\end{equation}
(see Figure 4h).
In the models below we adopt $\Delta t = 360$ s and
$\tau_{j} = 20$ s, which are consistent with the observations
of Berger et al.\  (1998).
The velocity amplitude of the jump $\sigma_j$ is known with
much less certainty because it represents the ``tail'' of the
observed distribution of speeds.
Berger \& Title (1996) found speeds up to 5 km s$^{-1}$, so
this seems to be a rough upper limit to $\sigma_j$.
This is our only true free parameter.

Because the walk and jump phases of MBP motion seem to be
statistically uncorrelated, we compute the full kinetic
energy spectrum $P_K$ as the simple sum of $P_{K,w}$ and
$P_{K,j}$ (see Figure 4i).
It is then possible to integrate this spectrum over frequency
(see eq.~[\ref{eq:UKearly}]) to obtain the transverse velocity
variance of MBP motions in the photosphere:
\begin{eqnarray}
  \langle \delta V \rangle^{2}_{\odot} & = &
    2 \sigma_{w}^{2} +
    \sqrt{2\pi} \, \sigma_{j}^{2} \tau_j / \Delta t \\
  & = & \left[ 1.6 \, + \, 0.139
  \left( \frac{\sigma_j}{1 \, \mbox{km} \, \mbox{s}^{-1}}
  \right)^{2} \right] \, \mbox{km}^{2} \, \mbox{s}^{-2}
  \label{eq:dvphot}
\end{eqnarray}
where the latter expression uses the values adopted above
for $\sigma_w$, $\tau_j$, and $\Delta t$.
Note that even for a large impulsive velocity of
$\sigma_{j} \approx 6$ km~s$^{-1}$, the root-mean-squared
velocity is significantly smaller than this value
($\langle \delta V \rangle_{\odot} \approx 2.5$ km~s$^{-1}$)
because the jumps occur infrequently.

Finally, we also compute the {\em total} energy spectrum
$P_{\rm tot}$ (i.e., including contributions from both kinetic
and magnetic energy) at the photosphere using the analytic
relations for isothermal thin flux tubes given in Appendix A
(specifically, eq.~[\ref{eq:isopart}]).
For very high frequencies the kinetic and magnetic energy
components are in equipartition, and $P_{\rm tot}$ is
just twice $P_K$.
For very low frequencies the kink-mode waves are evanescent
and the physically realistic solution contains much more
kinetic energy than magnetic energy (thus,
$P_{\rm tot} \approx P_{K}$).
The resulting total energy spectrum, plotted as a solid line
in Figure 4i, is essentially our lower boundary condition on
the amplitudes of Alfv\'{e}n waves of various frequencies.
We describe how this information is folded into the global
solutions in {\S}~5.3.

\section{Non-WKB Wave Analysis}

We model the transverse wave properties in the open magnetic
regions described in {\S}~3 as purely linear perturbations
to the assumed zero-order background plasma state.
The basic MHD equations that need to be solved are the
mass and momentum conservation equations and the magnetic
induction equation, given by
\begin{equation}
  \frac{\partial \rho}{\partial t} +
  \nabla \cdot (\rho {\bf v}) \, = \, 0
  \label{eq:mascon0}
\end{equation}
\begin{equation}
  \frac{\partial {\bf v}}{\partial t} +
  ({\bf v} \cdot \nabla) {\bf v} \, = \,
  -\frac{1}{\rho} \nabla p + {\bf g} + \frac{1}{4\pi\rho}
  \left[ ( \nabla \times {\bf B} ) \times {\bf B} \right]
  \label{eq:momcon0}
\end{equation}
\begin{equation}
  \frac{\partial {\bf B}}{\partial t} \, = \,
  \nabla \times \left( {\bf v} \times {\bf B} \right)
  \label{eq:induct0}
\end{equation}
where the velocity ${\bf v}$ and magnetic field ${\bf B}$
are not yet separated into zero-order and first-order parts,
$p$ is the gas pressure, and ${\bf g}$ is the gravitational
acceleration.

We apply these equations in two regions with very different
physics:
\begin{enumerate}
\item
{\em Below the merging height} ({\S}~5.1)
we model the waves as incompressible Lagrangian perturbations of
the central axis of a thin, strong-field flux tube that
expands superradially and is surrounded by a field-free region.
In this region we assume the outflow speed of the solar wind
is negligibly small.
In addition to the properties within the thin tube, we also
specify the density external to the tube $\rho_e$ in
the field-free region.
\item
{\em Above the merging height} ({\S}~5.2)
we model the waves as incompressible Eulerian perturbations
filling the volume of the expanding flux tube, which is assumed
to be surrounded by similar tubes.
All background properties, including a nonzero solar wind
speed, vary only in the radial direction.
\end{enumerate}
For both regions we transform the MHD equations above into
wave equations with an assumed $e^{i \omega t}$ time dependence.
The equations are then solved ``monochromatically'' for a grid
of frequencies, also assuming that in each solution the frequency
remains constant as a function of height.
We do not make the WKB (i.e., eikonal) approximation\footnote{%
By ``WKB'' we refer broadly to the use of an asymptotic
expansion that facilitates the solution of the linear
differential equations.  Specifically, for the wave equations
presented in this paper, the ``WKB limit'' is that of pure
outward propagation with no reflection.
The use of this acronym that cites the contributions of Wentzel,
Kramers, and Brillouin does not imply neglect of the earlier work
of Liouville, Green, Carlini, Rayleigh, Jeffries, and others.}
that wavelengths are small compared to background scale lengths;
indeed we do not even need to define the concept of wavelength
because the complete spatial oscillation pattern is computed
numerically as a function of height.
The solutions for individual frequencies are subsequently
assembled into a full radially varying power spectrum,
normalized by the empirically derived power spectrum at
the photosphere ({\S}~4).

The following subsections present the specific equations
that we solve in the two regions outlined above.

\subsection{Thin Flux Tubes}

Below the merging height, the MBP flux tubes are shaken
transversely and kink-mode waves are excited (see also
Wilson 1979; Spruit 1981; Ulmschneider et al.\  1991).
For incompressible perturbations about the equilibrium
state, the density $\rho$ is a zero-order quantity, the
velocity ${\bf v}$ is horizontal and a first-order quantity,
and the magnetic field ${\bf B}$ has a zero-order vertical
component and a first-order horizontal component.
Following the motion of the thin tubes, we write the
Lagrangian forms of the momentum and induction equations
as follows:
\begin{equation}
  \rho \frac{d {\bf v}}{dt} \, = \, \rho {\bf g}
  - \nabla \left( p + \frac{B^2}{8\pi} \right) +
  \frac{( {\bf B} \cdot \nabla ) {\bf B}}{4\pi}
  \label{eq:momconLag}
\end{equation}
\begin{equation}
  \frac{d {\bf B}}{dt} \, = \,
  \left( {\bf B} \cdot \nabla \right) {\bf v}
  \label{eq:inductLag}
\end{equation}
where the advective derivative
\begin{equation}
  \frac{d}{dt} \, = \,
  \frac{\partial}{\partial t} +
  {\bf v} \cdot \nabla
\end{equation}
follows the motion of the tube's central axis.
Above we have assumed that $\nabla \cdot {\bf v} = 0$ as
a statement of mass conservation for incompressible flows.

We write the scalar horizontal perturbations in velocity
and magnetic field as $v_{\perp}$ and $B_{\perp}$, and we
implicitly assume linear polarization of the waves
in a single transverse dimension.
In many ways, though, the equations to be derived are
degenerate with toroidal Alfv\'{e}n waves at larger heights
(e.g., Heinemann \& Olbert 1980) and our assumption should
not appreciably limit the generality of the results.
(Other polarization modes have been studied by, e.g.,
Spruit 1982, 1984; Lou 1993; Roberts 2000;
Noble, Musielak, \& Ulmschneider 2003; Ruderman 2003.)

In order to derive the wave equations we note three aspects
of thin tubes in the solar atmosphere:
\begin{enumerate}
\item
For an oscillating flux tube, the direction perpendicular
to its instantaneous axis will, generally, be inclined
with respect to the radial direction away from the Sun.
The components of the above equations parallel to the flux
tube axis are uninteresting, and will be considered to be
``solved'' by the zero-order background state described
in {\S}~3.
\item
We assume transverse total pressure balance between the
flux tube and the surrounding field-free region, and that the
field-free region is in simple hydrostatic equilibrium.  Thus,
\begin{equation}
  \nabla \left( p + \frac{B^2}{8\pi} \right) \, = \,
  \nabla p_{e} \, = \, \rho_{e} {\bf g}
  \,\,\, .
\end{equation}
\item
The motion of the tube induces motions in the surrounding
field-free region, which in turn must have a back-reaction on
the tube's original motion.
Spruit (1981) took this into account by increasing the apparent
inertia of the tube.
Thus, for the perpendicular component of eq.~(\ref{eq:momconLag}),
the factor of $\rho$ in the advection term on the left-hand side
must be replaced by $(\rho + \rho_{e})$.
This is equivalent to the assumption that the tube ``carries
along'' a parcel of the surrounding fluid with equal kinetic
energy density to that of the tube itself.
Osin, Volin, \& Ulmschneider (1999) reviewed different approaches
to the inclusion of this back-reaction effect and found that
Spruit's (1981) approach adequately describes the physics for
transverse oscillations of a nearly vertical flux tube.
\end{enumerate}

With the above considerations, Spruit (1981) showed that the
perpendicular component of eq.~(\ref{eq:momconLag}) can be
expressed as a linearized wave equation
\begin{equation}
  \frac{\partial^{2} v_{\perp}}{\partial t^2} \, = \,
  \frac{g \, \Delta\rho}{\rho_{\rm tot}}
  \, \frac{\partial v_{\perp}}{\partial r} \, + \,
  V_{\rm ph}^{2} \frac{\partial^{2} v_{\perp}}{\partial r^2}
  \label{eq:momSpruit}
\end{equation}
where the magnitude of the gravitational acceleration
$g = GM_{\odot}/r^2$ is nearly constant over the
heights we consider.
In Spruit's (1981) ideal limit that the flux tubes are
completely isolated from one another, the density
quantities introduced above are defined as
\begin{equation}
  \rho_{\rm tot} \, = \, \rho + \rho_{e}
\end{equation}
\begin{equation}
  \Delta \rho \, = \, \rho - \rho_{e}
\end{equation}
and $V_{\rm ph} \equiv B_{0} / \sqrt{4\pi \rho_{\rm tot}}$
is a modified kink-mode phase speed that takes the surrounding
inertia into account.
Note that in the Lagrangian picture, $v_{\perp}$ is the time
derivative of the horizontal displacement $\xi$.
The first term on the right-hand-side of eq.~(\ref{eq:momSpruit})
is due to the {\em buoyancy} of the low-density flux tube.
The second term on the right-hand-side is the magnetic tension
restoring force due to the curvature of the flux tube.
The Lagrangian induction equation is given simply by
\begin{equation}
  \frac{\partial B_{\perp}}{\partial t} \, = \,
  B_{0} \, \frac{\partial v_{\perp}}{\partial r}
  \,\,\, .
  \label{eq:indSpruit}
\end{equation}

Analytic solutions to the above equations are possible when
the radial derivative terms have constant coefficients;
this occurs in an exponential isothermal atmosphere (see
Appendix A for details).
Traditionally, the solutions to these equations display
evanescence for frequencies below a critical cutoff value
$\omega_c$.
For the adopted background state at the photosphere ($z=0$),
$\rho_{e} / \rho = 2.35$ and $V_{\rm ph} = 6.672$ km~s$^{-1}$,
and Appendix A gives an analytic estimate for the corresponding
critical period ($2\pi / \omega_c$) of about 12.5 minutes.
This period is significantly longer than the acoustic cutoff
frequency of 3--6 minutes in the photosphere and chromosphere.
Thus, the kink mode has been suspected for several decades as
being able to transport more convective wave energy up to
the corona than acoustic waves.

Before discussing our numerical solutions to
eqs.~(\ref{eq:momSpruit})--(\ref{eq:indSpruit}) between $z=0$
and $z_m$, one simplification assumed above must be reexamined.
Just below the merging height, the flux tubes cannot be
considered truly isolated from one another.
The enhanced inertia assumed by Spruit (1981) assumed that
the surrounding fluid carried along by a given tube is all
field-free, but near the merging height this is not the case.
Spruit (1982) gave the equations for selected kink-mode wave
properties for the general case where the surrounding medium
has a nonzero field strength, but here we deal with the
encroachment of neighboring flux tubes in a simpler manner.
In the above equations we express $\rho_{\rm tot}$ and
$\Delta \rho$ by
\begin{equation}
  \rho_{\rm tot} \, = \, \rho + (1 - \phi) \rho_{e}
\end{equation}
\begin{equation}
  \Delta \rho \, = \, (\rho - \rho_{e}) (1 - \phi)
  \label{eq:delrho}
\end{equation}
where $\phi$ is essentially a statistical filling factor
of neighbor tubes within the near-tube region that gets carried
along with a tube's oscillatory motion.
The isolated tube limit is $\phi = 0$, but at the merging height
(and above), $\phi = 1$.
The specific form of the modifications above were constrained by
the need for both $V_{\rm ph} = V_{A}$ and for the buoyancy term
in eq.~(\ref{eq:momSpruit}) to vanish in the ``merged''
limit of $\phi = 1$.
The reduction of $\Delta\rho$ also reduces the critical
frequency for evanescence (see eq.~[\ref{eq:isoomc}]), and at
the merging height $\omega_{c} \rightarrow 0$.
We derive $\phi (z)$ by using magnetic flux conservation
together with the assumption that the overall area subtended by
the full network patch is constant between the photosphere and
the merging height.
Thus,
\begin{equation}
  \phi (z) \, = \, \left\{
  \begin{array}{ll}
    B_{0} (z_{m}) / B_{0} (z) \,\,\, , & z < z_{m} \\
    1 \,\,\, , & z \geq z_{m} \\
  \end{array}
  \right.
  \label{eq:phidef}
\end{equation}
where $B_{0} (z_{m}) = 120.4$~G.
Because $B_0$ increases rapidly with decreasing height below
$z_m$, $\phi$ rapidly drops from 1 (at $z_{m} = 600$~km) to
0.5 at $z = 450$~km, then more slowly down to 0.08 at $z = 0$.
In Figure 3 we plot the resulting values of $\rho_{\rm tot}$
(as a hydrogen number density) and $V_{\rm ph}$ below the
merging height.

The wave equation (eq.~[\ref{eq:momSpruit}]) is simplified by
assuming an $e^{i \omega t}$ time dependence with a real
frequency $\omega$.
We solve numerically for the radial dependence of $v_{\perp}$
by expressing the second-order wave equation as two coupled
first-order ordinary differential equations in $v_{\perp}$ and
$\partial v_{\perp} / \partial r$ (with both quantities
assumed to be complex), and using fourth-order Runge-Kutta
integration (e.g., Press et al.\  1992).
The upper boundary conditions at $z_m$ are specified by the
solutions of the wave equations {\em above} $z_m$ (see next
section), and the numerical integration proceeds from $z_m$
down to the photosphere.
The coupling of solutions below and above the merging height
is discussed in {\S}~5.3.

\subsection{Wave Reflection in the Solar Wind}

Above the merging height, the transverse incompressible
fluctuations act as MHD Alfv\'{e}n waves, and our solution
procedure largely follows that of Heinemann \& Olbert (1980)
and Barkhudarov (1991).
Formally, the mechanisms of WKB theory can be extended to
describe linear wave reflection (e.g., Hollweg 1990), but we
follow the usual non-WKB formalism in order to solve for
the radial dependence of the transmitted and reflected wave
properties from the merging height all the way to a distance
of 4 AU.

The monochromatic non-WKB wave transport equations are derived
from the mass, momentum, and induction equations listed above
(eqs.~[\ref{eq:mascon0}]--[\ref{eq:induct0}]) in the limits
that all background quantities vary only in radius and that
the velocity and magnetic field perturbations are perpendicular
to the zero-order field direction.
Khabibrakhmanov \& Summers (1997) showed how to treat general
vector operations in a superradially expanding flux tube.
We express the wave properties in terms of Elsasser (1950)
variables, defined here as
\begin{equation}
  z_{\pm} \, \equiv \, v_{\perp} \pm
  \frac{B_{\perp}}{\sqrt{4 \pi \rho}}
  \label{eq:elsdefine}
\end{equation}
(see also Tu \& Marsch 1995),
with $z_{-}$ representing outward propagating waves and
$z_{+}$ representing inward propagating waves.
In terms of these variables, the incompressible first-order
equations are expressed as two coupled transport equations:
\begin{equation}
  \frac{\partial z_{\pm}}{\partial t} + ( u \mp V_{A} )
  \frac{\partial z_{\pm}}{\partial r} \, = \, ( u \pm V_{A} )
  \left( \frac{z_{\pm}}{4 H_D} + \frac{z_{\mp}}{2 H_A}
  \right)
  \label{eq:zpm}
\end{equation}
where the (signed) scale heights are
$H_{D} \equiv \rho / (\partial \rho / \partial r)$ and
$H_{A} \equiv V_{A} / (\partial V_{A} / \partial r)$.
These equations are valid for superradial divergence and for
all values of the zero-order outflow speed $u$.
Note that linear reflection arises because of the $z_{\mp}$
term on the right-hand side.

In Appendix B we discuss additional details about these equations
and how they are equivalent to other versions given in
earlier work.
To our knowledge, the ``compact'' form of eq.~(\ref{eq:zpm})
has not been recognized fully, although the expressions of
Heinemann \& Olbert (1980) and Khabibrakhmanov \& Summers (1997)
were closely related.
The above form of eq.~(\ref{eq:zpm}) is particularly useful in
showing the large-scale density dependence of the wave amplitude
in two limiting cases when the outward propagating waves are
dominant (i.e., $|z_{-}| \gg |z_{+}|$).
Near the Sun, where $u \ll V_A$, we can approximate the
lower-sign version of the above equation as
\begin{equation}
  \frac{\partial z_{-}}{\partial r} \, \approx \,
  -\frac{z_{-}}{4\rho} \frac{\partial \rho}{\partial r}
  \,\,\, ,
\end{equation}
and thus $z_{-}$ is proportional to $\rho^{-1/4}$ as predicted
by WKB theory (see also Moran 2001).
Similarly, far from the Sun, where $u \gg V_A$, $z_{-}$ is
seen to be proportional to $\rho^{+1/4}$.
More general properties of the non-WKB solutions of
eq.~(\ref{eq:zpm}) are discussed in {\S}~6, and also by
MacGregor \& Charbonneau (1994) and Krogulec et al.\  (1994).

As with the solutions below the merging height, we assume
an oscillatory time dependence of the Elsasser variables
(i.e., $e^{i \omega t}$ with a real, constant frequency),
and we solve for their radial dependence numerically.
We restrict our solutions to positive frequencies and note
that taking the negative of a given frequency produces
solutions to eq.~(\ref{eq:zpm}) that are the complex conjugates
of the analogous solutions obtained with $\omega > 0$.
Thus, the radial evolution of physical quantities (i.e.,
real wave amplitudes) is unaffected by the sign of $\omega$.
The existence of the Alfv\'{e}n critical point complicates
the numerical solution of eq.~(\ref{eq:zpm}), but we follow
the general solution procedure outlined by Barkhudarov (1991).
Once the oscillatory time dependence has been assumed, the
complex amplitudes $z_{+}$ and $z_{-}$ are expressed as the
products of real amplitudes and phase factors of unit magnitude.
There are then four ordinary differential equations for
these quantities that are solved first at the Alfv\'{e}n
critical point $r_A$ (analytically, using certain physicality
constraints such as the requirement that the outward wave
energy always exceed the inward wave energy), then we
integrate numerically using the fourth-order Runge-Kutta
method both upwards and downwards from $r_A$.
Some of Barkhudarov's (1991) expressions had to be modified
to take account of the superradial divergence of the magnetic
field.
The linear amplitudes $| z_{\pm} |$ are specified only
to within an arbitrary normalization factor, although
both the phase factors and all {\em relative} quantities
(such as ratios of Elsasser amplitudes at different radii)
do not depend on this normalization.

\subsection{Solution of the Coupled Wave Equations}

Our baseline model consists of a grid of 300 frequencies,
evenly spaced in $\log\omega$ over five orders of magnitude
with periods ranging from 3 to 300,000 s.
The discrete radial grid extends from just above the photosphere
($z = 10$ km) to 4 AU ($z = 859 \, R_{\odot}$) and contains
11809 grid points distributed mainly logarithmically, but with
some regions (like the transition region) sampled more finely.
In the photosphere, chromosphere, and low corona (below $x_0$),
the relative grid separation $\Delta z / z$ is 0.00064.
Within the most rapidly changing 20 km of the transition region,
$\Delta z / z$ is decreased to 0.0001.
In the extended corona and solar wind, $\Delta z / z$ is made
to gradually increase to 0.016 at the outer boundary.
The Runge-Kutta algorithm also has an adaptive stepsize that
subdivides the above grid zones until a relative accuracy of
$10^{-9}$ is achieved in the integration variables.
This degree of accuracy is needed to follow the oscillatory
behavior of the waves.

The non-WKB wave equations are solved first for each frequency
in the upper corona/wind region as described in {\S}~5.2, and
the resulting Elsasser variables at the lower boundary (i.e.,
the merging height) are used to compute the complex values of
$v_{\perp}$ and $B_{\perp}$ at that height.
These quantities, still with an arbitrary degree of normalization,
are used as the upper boundary conditions for the numerical
solution of the flux-tube wave equations given in {\S}~5.1.
The induction equation, eq.~(\ref{eq:indSpruit}), is used only
to convert the boundary condition for $B_{\perp}$ into a
condition for $\partial v_{\perp} / \partial r$.
We make the assumption that all of the Alfv\'{e}nic wave energy
in the upper region is converted smoothly into kink-mode wave
energy in the lower region.
After the transport equations are solved in both regions,
the photospheric MBP power spectrum (derived in {\S}~4)
is used to renormalize the wave power quantities at all heights.
To show how this is done we must first define the kinetic and
magnetic energy densities for each monochromatic model:
\begin{equation}
  E_{K} \, \equiv \, \frac{\rho \, v_{\perp}^{\ast} v_{\perp}}{2}
  \,\,\, , \,\,\,\,\,\,\,
  E_{B} \, \equiv \, \frac{B_{\perp}^{\ast} B_{\perp}}{8\pi}
  \label{eq:EKB}
\end{equation}
and also the supplementary quantities
\begin{equation}
  E_{\pm} \, \equiv \, \frac{\rho \, z_{\pm}^{\ast} z_{\pm}}{4}
  \,\,\, , \,\,\,\,\,\,\,
  E_{\times} \, \equiv \, \frac{\rho \, \mbox{Re} ( v_{\perp}^{\ast}
  B_{\perp} )} {\sqrt{4\pi\rho}} \, = \, E_{+} - E_{-}
  \label{eq:EPM}
\end{equation}
where below the merging height we use $\rho_{\rm tot}$ for the
above densities.
The energy densities defined above do not depend on time $t$
since they contain products of $e^{i \omega t}$ and its
complex conjugate.
These definitions also ensure that the total fluctuation energy
for each frequency satisfies
$E_{\rm tot} = E_{K} + E_{B} = E_{+} + E_{-}$.
In the simple WKB theory (i.e., all outward propagating waves),
$E_{+} = 0$, $E_{K} = E_{B}$, and
$E_{\rm tot} = E_{-} = -E_{\times}$, and the departure from
WKB theory can be assessed roughly by departures from these
ideal energy identities.

In {\S}~4 we derived the total power
$P_{\rm tot} (\omega, R_{\odot})$ at the photosphere.
The numerical integrations described above gave us the
various $E$ quantities on a two-dimensional grid in $\omega$
and radius (with the energies for each frequency known up to
an arbitrary multiplicative constant).
We thus compute ``renormalized'' power spectra on the
discrete grid as
\begin{equation}
  \left\{ \begin{array}{c}
    P_{K} (\omega,  r) \\
    P_{B} (\omega,  r) \\
    P_{\pm} (\omega,  r) \\
    P_{\rm tot} (\omega,  r)
  \end{array} \right\}
  \, = \, \frac{P_{\rm tot} (\omega, R_{\odot})}
  {E_{\rm tot} (\omega, R_{\odot})} \times
  \left\{ \begin{array}{c}
    E_{K} (\omega,  r) \\
    E_{B} (\omega,  r) \\
    E_{\pm} (\omega,  r) \\
    E_{\rm tot} (\omega,  r)
  \end{array} \right\}  \,\, .
  \label{eq:PEscale}
\end{equation}
Once the power spectra have been defined we can then
specify various frequency-averaged energy densities.
In general, we define
\begin{equation}
  U_{\rm tot} \, \equiv \,
  \int_{0}^{\infty} d\omega \, P_{\rm tot} (\omega)
\end{equation}
with analogous definitions for $U_K$, $U_B$, and $U_{\pm}$
(see also eq.~[\ref{eq:UKearly}]).
For ease of interpretation we also define the frequency-averaged
velocity, magnetic, and Elsasser amplitudes as the quantities
in angle brackets below:
\begin{equation}
  U_{K} \, = \, \frac{\rho \langle \delta V \rangle^2}{2}
  \,\,\, , \,\,\,\,\,\,\,
  U_{B} \, = \, \frac{\langle \delta B \rangle^2}{8\pi}
  \label{eq:UKUB}
\end{equation}
\begin{equation}
  U_{\pm} \, = \, \frac{\rho \langle Z_{\pm} \rangle^2}{4}
  \,\,\, .
  \label{eq:UPUM}
\end{equation}

\section{Results and Observational Implications}

\subsection{Linear Wave Properties}

The procedures outlined above resulted in a large amount of
numerical data ($\sim$350 megabytes) describing the behavior of
non-WKB kink-mode and Alfv\'{e}n waves as a function of
frequency and radius.
In this section we attempt to distill and present the salient
results in three gradual steps:
(1) in Figures 5--7 we present frequency-dependent wave properties
that have not yet been renormalized to the photospheric power
spectrum (and thus are plotted as dimensionless ratios),
(2) in Figure 8 we show the total power spectrum as a function of
frequency for selected radii, and
(3) in Figures 9--18 we present various frequency-integrated
quantities that depend on the photospheric power normalization.

In order to determine to what degree the waves in various
regions depart from ideal WKB theory, we show in Figure 5 the
outward propagating wave action flux $f^2$ (defined in
Appendix B) for a selection of periods, each normalized to
the value of $f^2$ in the photosphere.
For fluctuations having wavelengths so small that the local
plasma parameters are approximately constant over several
wavelengths, wave action is conserved and the quantity $f$
is constant.
Note from Figure 5 that in the extended corona and solar wind
(e.g., $z \gtrsim 1 \, R_{\odot}$) Alfv\'{e}n waves with
periods shorter than a few hours obey wave action conservation,
but for periods exceeding $\sim$10 hours this breaks down.
Note that for {\em all} computed periods the wave action is
not conserved below the transition region
($z \approx 0.003 \, R_{\odot}$); this region acts
as a sufficiently sharp ``barrier'' to induce significant
reflection in the chromosphere (see also Wentzel 1978;
Hollweg 1981; Campos \& Gil 1999).

In much of the previous work on non-WKB Alfv\'{e}n wave
reflection, the departures from WKB theory have been
characterized as a function of frequency.
For frequencies exceeding a critical value $\omega_A$,
the resulting wavelengths are so short that the wave
propagates as if it were in a homogeneous medium and there
is negligible reflection.
For frequencies lower than $\omega_A$ there is significant
reflection, and as $\omega \rightarrow 0$ the oscillation
approaches the properties of a standing wave with equal
inward and outward power.
In a magnetized hydrostatic atmosphere, the critical
frequency is given by the local value of
$| \partial V_{A} / \partial r |$, or more precisely for
arbitrary expansion factors,
$| \nabla \cdot {\bf V}_{A} |$
(e.g., Ferraro \& Plumpton 1958; An et al.\  1990).
In a supersonic wind, though, the radial dependence
of the Alfv\'{e}n speed is no longer the dominant factor in
determining how much reflection takes place.
At large distances from the Sun, Heinemann \& Olbert (1980)
and Barkhudarov (1991) showed that $\omega_A$ is given
approximately by $u_{\infty} / 2 r_A$, where $u_{\infty}$
is the asymptotic outflow speed and $r_A$ is the Alfv\'{e}n
radius.
(We use this expression as a fiducial definition of
$\omega_A$, but we also note that it neglects several
order-unity correction factors that depend on the flow tube
geometry.)
For the zero-order solar wind model defined in {\S}~3.2
we find an associated critical period $2\pi / \omega_A$ of
about 30 hours, or 1.25 days.

Figure 6 shows contours of the Alfv\'{e}n ratio---i.e., the
ratio of kinetic to magnetic energy density $E_{K} / E_{B}$
as a function of period and height.
For ideal MHD Alfv\'{e}n waves in a homogeneous medium this
ratio is 1 and the waves are in energy equipartition
(Wal\'{e}n 1944).
Figure 6 can be broken up into four ``quadrants'' that have
the following limiting properties:
\begin{enumerate}
\item
{\em Upper left:}
For long periods below the transition region, most of the wave
energy is kinetic with only a negligible magnetic energy
density.
This is consistent with the predictions of kink-mode wave
theory for the so-called ``shallow'' evanescent solution.
In Appendix A we discuss several reasons why the solar atmosphere
is suspected to naturally prefer this solution.
\item
{\em Lower left:}
For short periods below the transition region, the Alfv\'{e}n
ratio rapidly fluctuates above and below 1.
This occurs because the wavelengths are small compared to the
photospheric and chromospheric scale heights, but they are
large compared to the scale heights in the transition region.
Thus significant reflection occurred and there is a superposition
of upward and downward waves.
The fluctuations are well described by standing waves having
fixed nodes in $v_{\perp}$ and $B_{\perp}$ (Hollweg 1981, 1984).
If one could separate the solutions into the component upward
and downward waves, their individual Alfv\'{e}n ratios would
be $\sim$1 as predicted by the kink-mode theory in Appendix A.
(This can also be seen roughly by averaging over several nodes.)
\item
{\em Upper right:}
For long periods above the transition region, most of the
wave energy is magnetic as was found by other non-WKB solar wind
models (e.g., Heinemann \& Olbert 1980).
These long periods correspond to ``quasi-static'' motions
of the field lines and have a similarity to the motions invoked
in DC theories of coronal heating (e.g., Kuperus et al.\  1981;
van Ballegooijen 1986; Milano, G\'{o}mez, \& Martens 1997).
\item
{\em Lower right:}
For short periods above the transition region, the plasma
appears homogeneous to the relatively small-wavelength
fluctuations and the ideal MHD equipartition holds.
\end{enumerate}

Figure 7 shows contours of the ratio $E_{+} / E_{-}$ which can
be thought of as an effective local reflection coefficient.
The ideal WKB theory corresponds to a ratio of zero (i.e.,
all outward propagation).
There is significant reflection below the transition region,
with a maximum value of this ratio of 0.99825 at the photosphere
for a period of $\sim$40 minutes.
As discussed in {\S}~5.2, we chose regularity conditions at
the Alfv\'{e}n critical point that ensured more outward than
inward power at all heights, and thus $E_{+} / E_{-}$ must
always be less than one.
Below the transition region, the behavior of $E_{+} / E_{-}$
versus frequency resembles analytic solutions that take account
of the chromospheric stratification and the strong reflection
at the transition region (e.g., Hollweg 1978a;
Schwartz, Cally, \& Bel 1984).
Because of the finite size of the atmosphere below the transition
region, a mild resonance structure is seen in the amount of
reflected wave flux (not apparent in Figure 7).
The resonances are not as sharp as in the isothermal models of
Schwartz et al.\  (1984), though, because the nonzero temperature
gradient results in a ``smearing'' of the preferred resonance
frequencies.

Above the transition region, the solutions plotted in Figure 7
are very similar to those of Heinemann \& Olbert (1980) and
Barkhudarov (1991); there is significant reflection when
$\omega \lesssim \omega_A$ and much less reflection for higher
frequencies (lower periods).
For frequencies above $\omega_A$---which we see below encompasses
most of the dominant part of the power spectrum---we can use
the analytic regularity conditions presented in {\S}~4 of
Barkhudarov (1991) to estimate the ratio $E_{+} / E_{-}$ at
the Alfv\'{e}n critical point:
\begin{equation}
  \frac{E_{+} (\omega, r_{A})}{E_{-} (\omega, r_{A})}
  \, \approx \, \frac{1}{\omega^2} \left|
  \frac{\partial V_A}{\partial r} \right|_{r = r_A}^{2}
  \,\,\,\,\,\,\,\, \mbox{for} \,\, \omega \gg \omega_{A}
  \,\,\, .
\end{equation}
By comparing to the numerical results in Figure 7 we verify
that this relation is accurate for periods less than
$\sim$200 minutes.
It may be possible to also estimate the radial dependence
of this ratio using similar analytic formulae.
We defer this to future work, but note that this would be
useful to models of coronal heating via turbulent cascade
(see {\S}~6.2.2).

Figure 8 shows the result of convolving the above results with
the photospheric power spectrum derived in {\S}~4.
In this figure we plot the total power $P_{\rm tot} (\omega)$
for selected heights, and with a specific choice for the
parameter $\sigma_j$ of 3 km~s$^{-1}$ (see below).
The amount of power loss between the photosphere and the
merging height is consistent with the analytic predictions of
isothermal kink-mode theory.
For high frequencies (i.e., periods less than 3--5 minutes),
$P_{\rm tot}$ decreases from $z=0$ to $z_m$ by a factor of
15 to 20; this is approximately the level of the decrease of
the background magnetic field between those heights, implying
that the propagating kink-mode relation $\rho v_{\perp}^{2}
\propto B_0$ applies.
For low frequencies (presumably evanescent), $P_{\rm tot}$
decreases by a factor of 190, which is very nearly equal to
the drop in density from $z=0$ to $z_m$.
This is consistent with the prediction of nearly constant
$v_{\perp}$ for the shallow evanescent solution discussed
in Appendix A (and thus $\rho v_{\perp}^{2} \propto \rho$).
The somewhat irregular structure that develops in the spectrum
between the photosphere and the merging height (and is passively
advected outwards above $z_m$) is not numerical noise.
Because of the nonisothermal temperature structure of the low
chromosphere and the use of a radially varying filling factor
$\phi$, the properties of the waves below the merging height
depend on frequency in a more complicated way than in the
ideal case described in Appendix A.
If the upward and downward propagating waves had identical
strengths and dispersive properties, the standing-wave nodes
in $E_K$ and $E_B$ would cancel out exactly in $E_{\rm tot}$.
However, $E_{\rm tot}$ exhibits a weak nodal structure because
of incomplete cancellation and thus contributes to the
irregularity in the power spectrum (see also
Schwartz et al.\  1984).

As in Figure 4, we plot in Figure 8 the product
$\omega P_{\rm tot}$ versus wave period to more clearly show
the periods that make the greatest contribution to the overall
wave energy.
We quantify this concept by defining the averaged, or
first-moment frequency as
\begin{equation}
  \langle \omega \rangle \, \equiv \,
  \frac{1}{U_{\rm tot}}
  \int_{0}^{\infty} d\omega \, \omega \, P_{\rm tot} (\omega)
  \,\,\, .
  \label{eq:omavg}
\end{equation}
We plot this quantity versus height in Figure 14 below, but
here we just note that the effective period
$2\pi / \langle \omega \rangle$ is about 3.5 minutes in the
photosphere, then it decreases to about 1.8 minutes above
the merging height.
Periods of a few minutes are natural to expect in the
photosphere and chromosphere, and possibly in the corona
as well (e.g., Chashei et al.\  1999).
However, it is reasonable to ask if these periods are expected
to dominate in interplanetary space.
In~situ spacecraft generally measure fluctuations in velocity,
density, and the magnetic field with the most power at periods of
a few hours (e.g., Goldstein et al.\  1995).
However, a spacecraft sitting still in the ecliptic plane at
1 AU would see our model network flux tube rotate past in about
4 to 5 hours, with its component flux tubes (each originating
in a different MBP) rotating by in substantially less than
one hour.
Thus it is possible that if the in~situ power spectra actually
do sample a ``fossil'' spectrum from the Sun, the dominant
periods of order 1 hour could be due to the passage of many
uncorrelated flux bundles past the spacecraft, and not the
waves within any one flux bundle.
(The ideal test would be to see if a spacecraft corotating with
the solar rotation period measures a significantly different
fluctuation spectrum than has already been observed.)

Figure 9, a key condensation of results for this paper, plots
the frequency-integrated velocity amplitude
$\langle \delta V \rangle$ as a function of height for a
selection of $\sigma_j$ parameters and compares it to several
different measurements of wave amplitudes.
We discuss each set of measurements briefly, by number, below:
\begin{enumerate}
\item
The dotted line shows a best-fit height dependence for the
microturbulence needed to match photospheric and
chromospheric line widths in the semiempirical VAL/FAL 
models (E.\  Avrett, 2003, personal communication; see
also Fontenla et al.\  1993, 2002).
\item
The filled circles show similar ``nonthermal'' line-broadening
velocities measured (on the solar disk) in the transition
region and low corona by the SUMER (Solar Ultraviolet
Measurements of Emitted Radiation; Wilhelm et al.\  1995)
instrument on {\em SOHO}
(the {\em Solar and Heliospheric Observatory}).
The height of each point has been estimated by matching the
height-dependence of temperature in the VAL/FAL model
with the assumed formation temperatures of the ions that
correspond to each point (Chae et al.\  1998).
\item
The crosses show nonthermal velocities inferred by SUMER
measurements made above the solar limb (Banerjee et al.\  1998).
Off-limb Doppler broadening observations are better suited for
measuring the properties of transverse Alfv\'{e}n waves than
observations on the solar disk.
\item
The gray region shows lower and upper limits on the nonthermal
velocity (Esser et al.\  1999) as computed from off-limb
measurements made by the UVCS (Ultraviolet Coronagraph
Spectrometer) instrument on {\em SOHO}
(Kohl et al.\  1995, 1997).
\item
The stars show early measurements (Armstrong \& Woo 1981)
of the random wavelike component of the solar wind velocity
from interplanetary scintillation observations of radio
signals passing through the corona, with the advecting
diffraction pattern being measured by more than one receiver.
\item
The error bars show a more recent determination of velocity
fluctuations---specifically transverse to the radial
direction---from radio scintillations (Canals et al.\  2002)
in the fast solar wind.
\item
The {\em Helios} and {\em Ulysses} probes measured time-averaged
Elsasser amplitudes that we converted to a representative
velocity amplitude using eqs.~(\ref{eq:UKUB})--(\ref{eq:UPUM})
and assuming $U_{K} = U_{B}$ in the heliosphere.
The data from both spacecraft were summarized by
Bavassano, Pietropaolo, \& Bruno (2000).
\end{enumerate}
Measured line-of-sight (one-dimensional) velocities have been
multiplied by $\sqrt{2}$ to take account of equivalent
fluctuations in both transverse dimensions (e.g.,
eq.~[\ref{eq:Pxy}]).

Because measurements (1) and (2) above refer to motions mainly
in the radial direction, we do not expect them to correspond
to transverse kink-mode or Alfv\'{e}n waves; they are plotted
mainly for heuristic comparison.
Note, though, that in Figure 9 we also plot (for the
$\sigma_{j} = 3$ km~s$^{-1}$ case) a dashed curve that shows
the frequency-averaged amplitude of the magnetic
fluctuations in velocity units:
\begin{equation}
  \langle \delta V \rangle_{B} \, \equiv \,
  \frac{\langle \delta B \rangle}{\sqrt{4 \pi \rho}}
\end{equation}
which for ideal MHD equipartition would be exactly equal to
$\langle \delta V \rangle$.
It is possibly a coincidence that this quantity so closely
matches the Chae et al.\  (1998) observations, but this
agreement may contain information about the {\em mode coupling}
between transverse and longitudinal waves in the transition
region.

The three choices for the MBP-jump velocity amplitude
$\sigma_j$ used in Figure 9 are 0, 3, and 6 km~s$^{-1}$.
The middle value seems to best match the off-limb nonthermal
line broadening measurements, and we will use this as a
baseline value in most subsequent calculations and plots.
Note that the in~situ measurements fall well below all
of the reasonable choices for $\sigma_j$ (even the lower
limit assuming no jumps whatsoever) and they exhibit a
steeper gradient than the undamped models.
The heliospheric ``deficit'' of wave power, compared to
most prior assumptions about the wave power in the solar
atmosphere, is well known (see, e.g.,
Roberts 1989; Lou 1993; Mancuso \& Spangler 1999).
In {\S}~6.2.2 we propose a potential solution of these
discrepancies based on a particular theory of turbulent wave
damping in the solar wind.

Figure 10 shows additional details about the height
dependence of the frequency-integrated wave properties.
The ratio $U_{K} / U_{B}$ shows the region of strong
departure from energy equipartition around the
transition region, and that kinetic energy exceeds
magnetic energy by a slight amount even down to the
photosphere.
The dimensionless magnetic amplitude
$\langle \delta B \rangle / B_0$ is less then 1 over most of
the computed range of heights (thus justifying the linear
approximation), but exceeds 1 above
$r \approx 19 \, R_{\odot}$.
Nonlinear calculations (e.g., Lau \& Siregar 1996) find
that $\langle \delta B \rangle / B_0$ does not grow so high
in the heliosphere and may saturate at values close to 1.

Figure 11 shows the height dependence of the two Elsasser
amplitudes.
Above the transition region, the frequency-integrated
outward propagating amplitude $\langle Z_{-} \rangle$ is much
larger than the inward amplitude $\langle Z_{+} \rangle$.
Below the transition region, the amplitudes approach a
constant ratio
$\langle Z_{-} \rangle / \langle Z_{+} \rangle \approx 1.027$.
We also plot the height dependence of $Z_{-}$ that would be
expected for ideal WKB wave action conservation (i.e.,
$f = \mbox{constant}$) and we scale it to the computed value
of $\langle Z_{-} \rangle$ at 4 AU.
The WKB value departs slightly from the non-WKB value between
the transition region and a height of $\sim$0.01 $R_{\odot}$
and is substantially smaller below the transition region where
there is strong inward wave power.
The undamped linear curves all disagree markedly with the
in~situ measurements of the Elsasser amplitudes (see, however,
{\S}~6.2.2).

Lastly, in Figure 12 we plot the frequency-integrated energy
flux density $F$ for the baseline $\sigma_{j} = 3$ km s$^{-1}$
case.
Below the merging height this quantity applies to the energy
flux carried upward by the MBP kink modes, and above the merging
height it applies to the volume-filling Alfv\'{e}n waves.
The wave flux is defined generally as
\begin{equation}
  F \, \equiv \, u (U_{K} + 2 U_{B}) + V_{A} (U_{-} - U_{+})
  \,\, ,
  \label{eq:fluxdef}
\end{equation}
which is based on the monochromatic definition given by
Heinemann \& Olbert (1980).
For outward propagating waves obeying wave action conservation,
one can write $F = U_{\rm tot} (1.5 u + V_{A})$.
In the lowest layers of the atmosphere one can also ignore
the factors above that depend on $u$ and thus the net flux
depends mainly on the difference between $U_{-}$ and $U_{+}$.
Because of the strong reflection below the transition region
there is strong cancellation; i.e., the magnitude of the purely
outward wave flux $V_{A} U_{-}$ exceeds the net flux by a
factor of about 20.
This means that $\sim$95\% of the kink-mode wave energy below
the transition region is ``trapped,'' with only 5\% escaping.

Analysis of a series of models having a range of $\sigma_j$
values yields a potentially useful explicit expression for the
photospheric wave flux in the MBPs.
Recalling that the random-walk and impulsive parts of the
power spectrum are assumed to be linearly independent (e.g.,
eq.~[\ref{eq:dvphot}]), we have fit the numerically computed
photospheric fluxes with the following function:
\begin{equation}
  \frac{F_{\odot}}{10^{7} \, \mbox{erg} \, \mbox{cm}^{-2} \,
  \mbox{s}^{-1}} \, \approx \, 9.75 \, + \, 2.48
  \left( \frac{\sigma_j}{1 \, \mbox{km} \, \mbox{s}^{-1}}
  \right)^{2}
\end{equation}
with less than 0.1\% uncertainty in the fit (mainly due to
roundoff error in the above expression having only three
significant figures).
Thus, for the reasonable range of $\sigma_{j}$ between 0 and 6
km s$^{-1}$, $F_{\odot}$ ranges between about 10$^8$ and 10$^9$
erg cm$^{-2}$ s$^{-1}$ (see also Musielak \& Ulmschneider 2001).
Note that for any given value of $\sigma_j$ the jump-term
in the above expression has a greater relative contribution to
the total flux than the corresponding jump-term in 
eq.~(\ref{eq:dvphot}) has to the photospheric velocity variance.
This occurs because the bulk of the power in the jump motions
is at higher frequencies than the walk motions and is less
susceptible to evanescent decay.
The flux is a frequency-integrated quantity that incorporates
this aspect of the solutions, whereas the velocity variance
does not.

We also plot in Figure 12 several other flux quantities.
The ``granule-averaged'' wave flux is given by the quantity
$\phi F$ (for the flux tube filling factor $\phi$, see
eq.~[\ref{eq:phidef}]), and it differs from $F$ itself
only below the merging height.
The quantity $\phi F$ spreads the wave energy out evenly
within the modeled network patch, and thus is a more
relevant quantity to compare with observations and predictions
of wave fluxes that arise in the network from the convection.
Note that the photospheric value of $\phi F = 2.7 \times
10^{7}$ erg cm$^{-2}$ s$^{-1}$ is of similar magnitude as the
{\em acoustic} wave flux of $5 \times 10^{7}$ erg cm$^{-2}$
s$^{-1}$ predicted by Musielak et al.\  (1994).
We also plot a ``supergranule-averaged'' flux $\phi_{c} F$,
where $\phi_c$ is a filling factor for the large-scale
funnel/canopy magnetic structure shown in Figure 2.
(We define $\phi_c$ only below the top height [12 Mm] of
the magnetostatic grid, as the ratio of the field strength
at that top height [11.8 G] to the field strength along
the central axis of the flux tube at lower heights.)
The quantity $\phi_{c} F$ is what one would use in order
to compute the total wave power (in erg s$^{-1}$) integrated
over the entire coronal hole.

\subsection{Wave Dissipation}

In this section we discuss two separate mechanisms that have
been proposed to damp Alfv\'{e}n waves in the solar atmosphere
and solar wind: viscosity ({\S}~6.2.1) and MHD turbulent
cascade ({\S}~6.2.2).
For clarity we do not treat these mechanisms together,
though in a completely self-consistent model all damping
mechanisms should be included and allowed to interact
with one another.

\subsubsection{Linear Viscous Dissipation}

The first damping mechanisms proposed for MHD waves in the
solar corona were collisional in nature (Alfv\'{e}n 1947;
Osterbrock 1961).
In the high-density plasma near the Sun, MHD waves can be
damped by viscosity, thermal conductivity, electrical
resistivity (i.e., Joule/Ohmic heating), and
ion-neutral friction.
For Alfv\'{e}n and fast-mode waves propagating parallel to
the magnetic field, though, the dominant collisional dissipation
channel in the corona is believed to be proton viscosity.
Here we estimate the viscous damping expected for the
background plasma conditions described above and find it
to be negligible.

We compute a collisional damping length $L_c$ as a function
of height $z$, and there should be appreciable damping only
where $L_{c} \lesssim z$.
Generally, the product of the damping length and a linear
damping rate $\gamma$ (i.e., the imaginary part of the
frequency) is the wave group velocity $V_{\rm gr}$ in the
inertial frame, given here for dispersionless Alfv\'{e}n
waves as $u + V_{A}$.
For viscous damping, $\gamma = \nu_{p} k^{2}$, where
$\nu_p$ is the proton kinematic viscosity and $k$ is the
radial wavenumber (van de Hulst 1951; Osterbrock 1961).
Thus, for Alfv\'{e}n waves propagating upward and parallel to
the background magnetic field, this leads to the general
expression
\begin{equation}
  L_{c} \, \approx \,
  \frac{( u + V_{A} )^{3}}{\nu_{p} \omega^{2}}
\end{equation}
(Tu 1984; Whang 1980, 1997).
In a strongly collisional plasma the kinematic viscosity is
given by the classical Braginskii (1965) formalism
(see also Hollweg 1986a).
In the collisionless solar wind, though, there is no clear
prescription for the effective viscosity.
We thus define the proton kinematic viscosity
phenomenologically as
\begin{equation}
  \nu_{p} \, = \, w_{p}^{2} \tau_{\rm eff} \, = \,
  \frac{k_{\rm B} T_{p} \tau_{\rm eff}}{m_p}
\end{equation}
where $w_p$ is the proton most-probable speed, $k_B$ is
Boltzmann's constant, $T_p$ is the proton temperature, $m_p$
is the proton mass, and $\tau_{\rm eff}$ is an effective
viscous timescale.
We adopt a fiducial proton temperature law for this analysis
that comes from the semiempirical VAL/FAL model (used for the
magnetostatic flux tube model) below $x = 1.0172$,
and we use
\begin{equation}
  T_{p} \, = \, \frac{5 \times 10^{5} \, \mbox{K}}
  {0.2 + 0.02 x^{0.8} + 0.21 x^{-33}}
\end{equation}
above 1.0172 $R_{\odot}$.
This expression reasonably reproduces the proton temperatures
measured by {\em Helios} and {\em Ulysses} in the highest-speed
solar wind (Tu, Freeman \& Lopez 1989; Goldstein et al.\  1996)
and it has a peak in the extended corona of $\sim$2.2 MK in
general agreement with UVCS/{\em{SOHO}} \ion{H}{1} Ly$\alpha$
measurements (e.g., Kohl et al.\  1998; Esser et al.\  1999).

For strong collisional coupling, the effective viscous time
scale $\tau_{\rm eff}$ is given by the mean time between
collisions (Braginskii 1965),
\begin{equation}
  \tau_{\rm coll} \, = \, \frac{3}{4} \sqrt{\frac{m_p}{\pi}}
  \frac{(k_{\rm B} T_{p})^{3/2}}{n_{\rm H} \, e^{4} \ln\Lambda_c}
\end{equation}
where $e$ is the proton/electron charge and $\ln \Lambda_c$
is the Coulomb logarithm, taken here to be 21.
The above timescale applies to the viscous damping of motions
{\em along} the magnetic field.
For shear motions transverse to the field, though, the
appropriate viscous time (even in the limit of strong
collisions) is reduced.
The off-diagonal Braginskii coefficients in the stress tensor
are obtained approximately by dividing $\tau_{\rm coll}$ by
dimensionless factors that take account of the magnetic field.
Thus,
\begin{equation}
  \tau_{\rm eff} \, \approx \, \frac{\tau_{\rm coll}}
  {(\Omega_{p} \tau_{\rm coll})^{m}} \,\, ,
\end{equation}
where $m$ is either 1 or 2, and $\Omega_p$ is the proton Larmor
frequency $e B_{0}/m_{p} c$.
The $m=0,1,2$ terms are roughly analogous to the direct, Hall,
and Pedersen components of the MHD conductivity in an ionized
plasma.

In the low-density collisionless limit, the classical $m=0$
case cannot apply because it would imply the viscosity
becomes infinitely large as the mean time between collisions
becomes infinite (the ``molasses'' limit).
Williams (1995) argued essentially that the general viscosity
in a collisional or collisionless plasma is found by taking
the shorter of either the $m=0$ or the $m=1$ timescales.
We assert, though, that the $m=2$ case seems to be the
most appropriate for the viscous damping of transverse
Alfv\'{e}n waves in the collisionless solar wind.
Consider the viscosity as an effective diffusion coefficient
($\ell^{2} / t$) describing scattering events with
mean-free-path length and time scales of $\ell$ and $t$,
respectively.
The actual energy losses arise from the interparticle collisions
with timescale $t \approx \tau_{\rm coll}$, but spatially the
{\em transverse} structure would be dominated by features with
sizes of order the proton thermal Larmor radius, and thus
$\ell \approx w_{p} / \Omega_{p}$.
For these values the viscosity reproduces the $m=2$ case.

Figure 13 shows the radial dependence of the linear damping
length $L_c$ for the three cases, $m=0$, 1, and 2, computed for
a wave period of 1 minute.
For heights below about 0.3 $R_{\odot}$ above the photosphere,
it is unlikely that any damping would take place because all
three cases have damping lengths much longer than the local
height.
This mid-corona distance ($r \approx 1.3$ to 2 $R_{\odot}$)
is approximately where collisions start to become unimportant
in coupling together electron, proton, and heavy ion plasma
properties (e.g., Cranmer, Field, \& Kohl 1999).
Above this height, then, the appropriate damping length should
transition to either the $m=1$ or the $m=2$ case, both of which
are substantially longer than the local height in the corona.
In the far solar wind, we believe the $m=2$ case is the most
realistic, and it always remains several orders of magnitude
larger than the local height.
Thus, our preliminary conclusion is that for both waves near
the peak of the power spectrum (periods of a one to a few
minutes), and for longer periods, linear viscous damping
is {\em unimportant} as a significant attenuation mechanism
for Alfv\'{e}n waves in the corona and fast solar wind.
For periods much shorter than 1 minute, though, this
kind of damping could be important---and may have already
been responsible for the sharp drop in power inferred from
photospheric MBP motions between 0.1 and 1 minute.

\subsubsection{Nonlinear Turbulent Dissipation}

The second type of wave damping we consider is turbulent
dissipation; i.e., a nonlinear cascade of energy from large
to small scales that terminates in an irreversible conversion
of wave energy into heat.
The need to include some kind of nonlinear damping or
saturation can be seen in Figure 10, where above a distance of
$\sim$20 $R_{\odot}$ the magnetic fluctuation amplitude begins
to exceed the background field strength (in opposition to
in~situ observations) and the linear assumption breaks down.
Addressing the full problem of anisotropic MHD turbulence in the
solar corona and solar wind, though, is well beyond the scope
of this paper (see, though, Cranmer \& van Ballegooijen 2003
for a summary of many related issues).
Here we include only one specific aspect of this theory:
a phenomenological damping rate that depends on the
properties of the largest scales in the turbulence and not
on the details of the cascade.

The presence of both outward and inward propagating
Alfv\'{e}n waves is a necessary prerequisite for the
nonlinear coupling to higher wavenumber that drives the
cascade (see, e.g., Matthaeus et al.\  1999;
Dmitruk, Milano, \& Matthaeus 2001).
Any expression for the turbulent energy decay must then
depend on both the $z_{-}$ and $z_{+}$ Elsasser amplitudes,
and thus the inclusion of turbulence introduces further
coupling between these two modes.
A straightforward phenomenological form for the nonlinear
transport has been suggested by Zhou \& Matthaeus (1990)
from the standpoint of ``reduced MHD'' (RMHD; see also
Strauss 1976; Montgomery 1982).
Effectively, the transport equation is the same as
eq.~(\ref{eq:zpm}), but with the following term
added to the right-hand side:
\begin{equation}
  - \frac{z_{\pm} | z_{\mp} |}{2 L_{\perp}}
\end{equation}
where $L_{\perp}$ is a transverse length scale representing
an effective correlation length of the turbulence---i.e.,
a similarity length scale associated with the energy transport
from large to small eddies.
(It may be imprecise to refer to this length scale as an
``outer scale,'' but for simplicity we treat this as a
synonym.)
We assume that $L_{\perp}$ scales with the transverse width of
the open flux tube; i.e., that it remains proportional to
$B_{0}^{-1/2}$.
Specifically, we normalize $L_{\perp}$ by specifying its
value at the merging height $z_m$.
We also define the constant $\Lambda$ as the ratio
of $L_{\perp}$ at the merging height to the transverse
radius of the network flux bundle at that height: 3 Mm
(see Figure 2).
Equivalently, we can express
\begin{equation}
  L_{\perp}(r) \, \approx \,
  \frac{33 \, \Lambda}{\sqrt{B_{0}(r)}} \,\, \mbox{Mm}
\end{equation}
if $B_0$ is measured in Gauss (see also Hollweg 1986b).

In order to know how best to implement the nonlinear term given
above, we need to understand over what range of heights this
term acts to appreciably damp the waves.
We do not expect a strong RMHD cascade to develop in the region
below the merging height because---despite the stochastic nature
of the fluctuation spectrum---the flux tubes do not strongly
interact with one another.
For heights above $z_m$ we can compare the effective damping
length $L_{\perp}$ to the local height $z$ above the photosphere.
For $\Lambda = 1$, we find that $L_{\perp} < z$ only above a
height of $z \approx 0.01 \, R_{\odot}$.
It makes sense that the chromosphere and transition region are
not expected to undergo much damping from this RMHD mechanism;
there seems to be insufficient time for the turbulent cascade
to develop in such small volumes.
Dmitruk \& Matthaeus (2003) discussed a hierarchy
of timescales that should be satisfied in order for this
turbulent damping process to be strong.
A simplified version of this hierarchy is as follows:
\begin{equation}
  t_{0} \, < \, t_{R} \, < \, t_{f}  \,\,\, ,
  \label{eq:tineq}
\end{equation}
where $t_0$ is a nonlinear outer-scale eddy cascade time,
$t_R$ is a timescale for Alfv\'{e}n wave reflection, and
$t_f$ is representative of the main driving period of the
waves.
In Figure 14 we compare several of these quantities to
see where these conditions are met.

In Figure 14 we plot a representative reflection time $t_R$
which we define as $1 / |\nabla \cdot {\bf V}_{A}|$.
This is probably of the same order of magnitude as the
Dmitruk \& Matthaeus (2003) ``Alfv\'{e}n wave crossing time''
$t_A$, since the latter can be defined as $H_{A} / V_{A}$,
or the wave travel time over a representative
Alfv\'{e}n-speed scale height $H_A$.
Thus, $t_{R} \approx t_{A}$.
For reference we also show the total wave travel time from
the photosphere to a given height, defined as
\begin{equation}
  t_{\rm trav}(r) \, \equiv \, \int_{R_{\odot}}^{r}
  \frac{dr'}{u(r') + V_{\rm ph}(r')}
  \,\,\, .
\end{equation}
We equate the driving period $t_f$ to the spectrum-integrated
first-moment period $2\pi / \langle \omega \rangle$
(see eq.~[\ref{eq:omavg}]) and we plot its value
for the baseline $\sigma_{j} = 3$ km s$^{-1}$ model.
Interestingly, $t_{f} > t_{R}$ only {\em below}
$z \approx 0.5 \, R_{\odot}$, and the hierarchy is not
satisfied at large distances where the damping length is
relatively small.
Note, though, that the demand in the above hierarchy for
$t_f$ to be the largest of the three timescales comes only
from the well-known non-WKB result that long periods are
the easiest to reflect, and thus they provide the most
turbulent mixing between inward and outward modes.
However, the amount of reflection in our model is a known
quantity (see, e.g., Figures 7 and 11), and it is irrelevant
whether it could have been maximized with longer-period waves.
Therefore we do not need to consider the last inequality in
eq.~(\ref{eq:tineq}) as a precondition for turbulent damping.

The key piece of the Dmitruk \& Matthaeus (2003) timescale
hierarchy is that the nonlinear driving time $t_0$ must be
short compared to the reflection time $t_R$.
Only when this condition is satisfied can the turbulent
cascade develop to the point where the wave damping can
occur efficiently.
In Appendix C we describe one way of using a model for
the anisotropic turbulent power spectrum to derive $t_0$
(see also Oughton, Dmitruk, \& Matthaeus 2004).
This timescale depends inversely on the outer-scale
length $L_{\perp}$, and in Figure 14 we plot the radial
dependence of $t_0$ for two extreme values of the
normalizing constant $\Lambda$.
An upper limit to the outer-scale length is given by
$\Lambda = 1$ (i.e., that the ``stirring'' takes place
on the spatial scale of the entire network element).
A reasonable lower limit for $L_{\perp}$ at the merging
height is the radius of the smallest expected MBP
flux tube.
At the merging height, the flux tubes have expanded to fill
the volume of the network patch, and thus the MBP radius
is given by half the nearest-neighbor distance $d_{nn}$.
For the lower-limit value of $d_{nn} \approx 350$ km
(see {\S}~3.1) we obtain a lower limit of $\Lambda = 0.06$.

We find in Figure 14 that $t_R$ begins to exceed $t_0$
somewhere above a height of 0.1 to 1 $R_{\odot}$.
We thus solve the nonlinearly modified wave transport
equation only above 0.1 $R_{\odot}$.
This restriction allows us to make a useful simplifying
assumption.
As seen in Figure 11, above this height there is only
relatively weak inward wave power; i.e.,
$\langle Z_{-} \rangle \gg \langle Z_{+} \rangle$.
We solve a modified time-steady wave action conservation
equation in the limit of pure outward propagation.
Note also that turbulence does not act in a straightforward
``monochromatic'' way, as was assumed in the constant-frequency
linear solutions to eq.~(\ref{eq:zpm}).
We thus ignore any frequency (and wavenumber) transport and
compute the overall effects of the turbulent damping
on the frequency-integrated wave energy (see, however,
Verdini \& Velli 2003).
Thus, we solve the following modified spectrum-averaged wave
action conservation equation in the limits of no time
dependence and dominant outward propagation:
\begin{equation}
  \frac{u V_A}{u + V_A} \frac{\partial}{\partial r}
  \left[ \frac{(u + V_{A})^{2} \langle Z_{-} \rangle^2}{u V_A}
  \right] \, = \, - \frac{\langle Z_{-} \rangle^{2}
  \langle Z_{+} \rangle}{L_{\perp}}
  \label{eq:zpmdamp0}
\end{equation}
(see Appendix B).
We use the linear solution for $\langle Z_{+} \rangle$ obtained
above in order to solve for the nonlinearly damped value of
$\langle Z_{-} \rangle$:
\begin{equation}
  \frac{\langle Z_{-} \rangle^2}{\langle Z_{-} \rangle_{0}^2}
  \, = \,
  \frac{(u_{0} + V_{A,0})^{2} \, u V_A}
  {(u + V_{A})^{2} \, u_{0} V_{A,0}} \,
  \exp \left\{ - \int_{r_0}^{r}
  \frac{dr' \, \langle Z_{+} (r') \rangle}
  {[u(r') + V_{A}(r')] L_{\perp} (r')} \right\}  \,\, ,
  \label{eq:zpmdamp}
\end{equation}
where all quantities with subscript `0' are given at the
effective lower boundary $z = 0.1 \, R_{\odot}$.
This is not a completely self-consistent method of obtaining
the damped wave power, since no back-reaction on
$\langle Z_{+} \rangle$ is computed, but it gives an adequate
order-of-magnitude result for the total power in the regions
dominated by outward propagating waves.

Figure 15 shows the result of integrating eq.~(\ref{eq:zpmdamp})
for three choices of $\Lambda$ (0.1, 0.35, and 1), as well as
the solution for no damping ($\Lambda \rightarrow \infty$).
We plot these solutions as velocity amplitudes in order to
facilitate comparison with the observational data, which are
also plotted as in Figure 9.
For each value of $\Lambda$, the resulting damped $U_{-}$
energy density was added to $U_{+}$ (which remains undamped)
to obtain $U_{\rm tot}$, and this total energy density was split
into kinetic and magnetic parts assuming that the linear undamped
ratio $U_{K}/U_{B}$ remains the same for the damped models.
The resulting $U_K$ was then used to compute
$\langle \delta V \rangle$ as in eq.~(\ref{eq:UKUB}).

The value of $\Lambda$ that produces the best agreement with
the in~situ measurements is approximately 0.35, implying a
transverse outer scale at the merging height of about 1100 km.
This value seems appropriate and consistent for motions excited
between granules of the same spatial scale.
If $\Lambda$ were much smaller than this value, it would imply
that the turbulence was dominated by internal motions in the
MBP flux tubes, which we essentially ignore in the thin-tube
models.
Furthermore, note from Figure 15 that for $\Lambda \leq 0.1$
there is so much damping that the approximation $\langle Z_{-}
\rangle \gg \langle Z_{+} \rangle$ breaks down above
$\sim$100 $R_{\odot}$ and we would have needed to solve a more
complicated set of damping equations.
The observational constraints which imply $\Lambda = 0.35$
allow us to avoid both of the above consistency problems that
arise for smaller values.

Properties of the $\Lambda = 0.35$ solutions were also plotted in
several other figures in order to compare with the linear results.
Figure 10 shows that the damped magnetic amplitude ratio
$\langle \delta B \rangle / B_0$ is smaller in interplanetary
space than the undamped ratio.
For the damped model, $\langle \delta B \rangle / B_0$ exceeds
1 only above 135 $R_{\odot}$ and does not exceed a value of
1.7 within the computed range of distances (see also
Lau \& Sireger 1996).
Figure 11 shows the better agreement between the damped
Elsasser amplitude $\langle Z_{-} \rangle$ and the measured
outward Elsasser energy.
The inward amplitude $\langle Z_{+} \rangle$ still does not
agree with the measured inward Elsasser energy (see also
Hollweg 1990), but note that the linear value was not modified
in the damping calculations done here.
Full simulations of the turbulence would probably see an
enhancement of $\langle Z_{+} \rangle$ and possibly even an
approach to inward/outward energy equipartition (e.g.,
Oughton \& Matthaeus 1995).

\subsection{Impact on the Steady-State Plasma}

A self-consistent treatment of waves in the solar atmosphere
and solar wind would consider the impact of fluctuations on
the mean, time-steady plasma properties.
Incorporating this back-reaction into the model steady state
is beyond the scope of this paper, but below we compute
a subset of the necessary source terms (e.g., the heating rate
and the wave pressure acceleration) for use in future models.

\subsubsection{Heating due to Turbulent Dissipation}

The nonlinear wave damping discussed in {\S}~6.2.2 produces
gradual heat deposition along an open magnetic flux tube.
The rate of energy conversion is derivable essentially
from eq.~(\ref{eq:zpmdamp0}), and here we give a slightly more
general version of the energy conversion rate based on a
comparison between phenomenological turbulence models and
numerical simulations.
The time rate of change in the general Elsasser variance
$\langle Z_{\pm} \rangle^2$ is given by
\begin{equation}
  \frac{d \langle Z_{\pm} \rangle^2}{dt} \, = \, -\alpha_{\pm}
  \frac{\langle Z_{\pm} \rangle^{2} \langle Z_{\mp} \rangle}
  {L_{\perp}}
  \label{eq:hossain}
\end{equation}
(e.g., Hossain et al.\  1995; Matthaeus et al.\  1999;
Dmitruk et al. 2001, 2002).
Comparison with numerical simulations found that the
dimensionless parameter $\alpha_{\pm}$ is of order unity.
We thus assume $\alpha_{\pm} = 1$ and define the total
heating rate $Q$ as
\begin{equation}
  Q \, \equiv \, -\frac{d U_{\rm tot}}{dt} = \, \rho \,
  \frac{\langle Z_{-}\rangle^{2} \langle Z_{+}\rangle +
        \langle Z_{+}\rangle^{2} \langle Z_{-}\rangle}
  {4 L_{\perp}}  \,\,\, .
  \label{eq:dmit}
\end{equation}
In our models this quantity is defined consistently only
above $z = 0.1 \, R_{\odot}$, where the damping is applied,
but below we calculate $Q$ for all heights along the flux
tube.
We caution that the heating rates below 0.1 $R_{\odot}$
are only estimates based on the undamped Elsasser amplitudes
at those heights.

Figure 16 plots the heating rate per unit mass ($Q/\rho$)
for the range of $\Lambda$ values shown in Figure 15
in addition to an even weaker-damping case of $\Lambda = 3$.
For the regions where the nonlinear damping does not change
the radial profiles of $\langle Z_{\pm} \rangle$ from their
undamped values, $Q$ scales with $\Lambda^{-1}$ as expected.
However, for heights where damping occurs the heating rate
is lowered because of the cumulative loss of wave power up to
that height.
Thus, for each height, there is a maximum heating rate that
defines a critical value $\Lambda_{\rm crit}$.
For $\Lambda > \Lambda_{\rm crit}$ the cumulative damping grows
weaker, and the decrease in $Q$ comes from the increase in the
denominator of eq.~(\ref{eq:dmit}).
For $\Lambda < \Lambda_{\rm crit}$, there is more cumulative
damping and the decrease in $Q$ comes from the loss of local
wave power in the numerator of eq.~(\ref{eq:dmit}).
For a distance of $r = 2 \, R_{\odot}$ in the extended corona,
$\Lambda_{\rm crit}$ is about 0.07, and the maximum heating
rate per unit mass at that value is
$\sim 3.9 \times 10^{11}$ erg s$^{-1}$ g$^{-1}$.

We also plot in Figure 16 an empirical heating rate (summed
from proton and electron contributions) that was derived in
order to reproduce a large number of remote-sensing and
in~situ measurements of fast solar wind conditions.
We use the heating rates from the SW2 model of Allen, Habbal,
\& Hu (1998).
For comparison with the above numbers, the empirically
constrained heating rate at $r = 2 \, R_{\odot}$ is about
$1.2 \times 10^{11}$ erg s$^{-1}$ g$^{-1}$
(see also Dmitruk et al.\  2002; Cranmer 2002).
In the extended corona ($r \approx 2$--3 $R_{\odot}$),
Figure 16 shows that the empirical heating rate agrees best
with the $\Lambda = 0.35$ case that also produces the right
amount of in~situ damping (Figure 15).
This rough agreement provides additional support for the
outer-scale normalization specified by this value.

It is worthwhile also to compare the heating rates computed
above with those computed using a damping rate appropriate
for isotropic Kolmogorov (1941) turbulence (see, e.g.,
Hollweg 1986b; Chae et al.\  1998; Li et al.\  1999;
Chen \& Hu 2001).
In order to derive a Kolmogorov heating rate appropriate
for quantitative comparison with the above results, we make both
the quasi-isotropic approximation $U_{-} = U_{+}$ and we assume
MHD energy equipartition $U_{K} = U_{B}$.
Thus, eq.~(\ref{eq:dmit}) becomes
\begin{equation}
  Q_{\rm kol} \, = \,
  \frac{\rho \, \langle \delta V \rangle^{3} \, \sqrt{2}}
  {L_{\perp}}
  \label{eq:kolm}
\end{equation}
where we use the damped velocity amplitude
$\langle \delta V \rangle$ as plotted in Figure 15.
The above expression differs by a factor of $\sqrt{2}$
from the simpler formulae given in many other papers;
this factor is required in order to remain consistent with
the isotropic/equipartition limit of eq.~(\ref{eq:dmit}).
Removing this factor would be equivalent to specifying
$\alpha_{\pm} = 1/\sqrt{2}$ in eq.~(\ref{eq:hossain}),
but we note that the {\em relative} comparison between
eqs.~(\ref{eq:dmit}) and (\ref{eq:kolm}) does not depend on
the value of $\alpha_{\pm}$.

In Figure 17 we plot the comparison between the complete
heating rate $Q$ and the Kolmogorov approximation $Q_{\rm kol}$.
For simplicity we show only the case having the preferred
$\Lambda = 0.35$ outer-scale normalization.
The curves are substantially different from one another
nearly everywhere, which indicates that the inward/outward
imbalance generated by non-WKB reflection is probably a
very important ingredient in Alfv\'{e}n wave heating models
of the solar wind.
The differences are small in the photosphere and low
chromosphere, where strong reflection leads to nearly equal
inward and outward wave power.
In the extended corona, though, the Kolmogorov heating rate
begins to exceed the anisotropic turbulent heating rate by as
much as a factor of 30.
The isotropic Kolmogorov assumption assumes the maximal
amount of possible mixing between inward and outward modes,
which is inconsistent with the relatively weak reflection
computed for the corona in our models.

Figure 17 also shows the radial dependence of an analytic
scaling relation for the turbulent heating rate given by
Dmitruk et al.\  (2002) for the region above the coronal
maximum in $V_A$.
This scaling relation gives
$Q \propto B_{0} | \partial V_{A} / \partial r |$,
and we show it for an arbitrary normalization.
The analytic heating rate per unit mass has a radial gradient
less steep than the other curves in Figure 17, but it has a
general shape that is similar.

\subsubsection{Wave Pressure Acceleration}

Waves that propagate radially through an inhomogeneous medium
exert a nondissipative net force on the gas.
This net momentum deposition has been studied for several
decades for both acoustic and MHD waves and is generally
called either ``wave pressure'' or the ponderomotive force
(e.g., Bretherton and Garrett 1968; Dewar 1970; Belcher 1971;
Alazraki and Couturier 1971; Jacques 1977).
Initial computations of the net work done on the bulk
fluid have been augmented by calculations of the acceleration
imparted to individual ion species (e.g.,
Isenberg \& Hollweg 1982; McKenzie 1994; Li et al.\  1999),
estimates of the departures from Maxwellian velocity
distributions induced by the waves (Goodrich 1978;
Hollweg 1978b), and extensions to nonlinearly steepened wave
trains (Gail, Ulmschneider, \& Cuntz 1990; Pijpers 1995).

For non-WKB Alfv\'{e}n waves propagating in a radial flux tube,
Heinemann \& Olbert (1980) gave the general second-order
expression for the wave pressure acceleration:
\begin{equation}
  {\bf a}_{\rm wp} \, = \,
  - ( {\bf v}_{\perp} \cdot \nabla ) {\bf v}_{\perp} +
  \frac{1}{4\pi\rho} \left[ ( \nabla \times {\bf B}_{\perp} )
  \times {\bf B}_{\perp} \right]
\end{equation}
which has the radial component
\begin{equation}
  a_{\rm wp} \, = \,
  - \frac{1}{8\pi\rho}
  \frac{\partial | B_{\perp} |^2}{\partial r} \, + \,
  \left( \frac{| B_{\perp} |^2}{8\pi\rho} -
  \frac{| v_{\perp} |^2}{2} \right)
  \frac{\partial}{\partial r} ( \ln B_{0} )  \,\, .
\end{equation}
In the limit of equipartition between the kinetic and magnetic
wave energy densities, the term in parentheses is zero.
For a spherical geometry (e.g., MacGregor \& Charbonneau 1994),
we note that the quantity $\partial ( \ln B_{0} ) / \partial r$
can be written simply as $-2/r$.

In our non-WKB model we have computed the frequency-integrated
kinetic and magnetic energy densities $U_K$ and $U_B$.
We find that the spectrum-weighted wave pressure acceleration
can be expressed in terms of these variables as
\begin{equation}
  \rho \langle a_{\rm wp} \rangle \, = \,
  - \frac{\partial U_B}{\partial r} +
  \left( U_{B} - U_{K} \right)
  \frac{\partial}{\partial r} ( \ln B_{0} )
  \,\,\, .
  \label{eq:awpavg}
\end{equation}
Figure 18 compares the computed values of
$\langle a_{\rm wp} \rangle$ for the undamped
($\sigma_{j} = 3$ km s$^{-1}$) and damped ($\Lambda = 0.35$) cases,
as well as an effective WKB wave pressure computed using just
the first term on the right-hand side of eq.~(\ref{eq:awpavg}).
There is some numerical noise in the plots that comes from the
necessity to take numerical derivatives.
Departures from the idealized WKB form arise only below
$z \approx 0.03 \, R_{\odot}$ and thus may not be important
for the acceleration of the solar wind.
In the photosphere, though, the non-WKB wave pressure acceleration
exceeds the approximate WKB acceleration by a factor of 100.
The WKB approximation also gives an unrealistically large spike
in $\langle a_{\rm wp} \rangle$ at the transition region
that does not appear in the exact non-WKB solution.
It remains to be seen if these differences would significantly
affect various observational constraints placed on the
properties of the transition region (e.g.,
Woods, Holzer, \& MacGregor 1990a, 1990b).

Figure 18 also shows the magnitude of the Sun's gravitational
acceleration $g$ as a function of height.
Presumably, the wave pressure acceleration has little net
effect on the corona and solar wind in regions where
$|\langle a_{\rm wp} \rangle | \ll g$.
For closed coronal loops, though, Laming (2004) computed
a similar non-WKB wave pressure acceleration and found that
its strength can rival that of gravity even in the chromosphere.
The Alfv\'{e}n wave reflection in these models is strong
enough to produce ion abundance variations that may explain
the observed dependence on First Ionization Potential (FIP).

\section{Summary of Major Results}

The goal of this work was to produce a detailed and
self-consistent description of the global energy budget of
transverse incompressible waves in open magnetic regions of
the solar atmosphere.
Here we list the unique features of the model, key results,
and some of the insights gained about the overall wave-plasma
system.
\begin{enumerate}
\item
Measurements of G-band bright points in the photosphere were used
to set the power spectrum of transverse waves as the lower
boundary condition of our model.
The observationally inferred power spectrum was summed from two
phases of MBP motion assumed to be statistically independent:
isolated random walks and occasional rapid jumps due to MBP
merging, fragmenting, or magnetic reconnection.
\item
The steady-state plasma density, magnetic field, and flow
velocity were constrained empirically from the photosphere to
a distance of 4 AU.
The successive merging of flux tubes on granular and supergranular
scales in the atmosphere was described using a two-dimensional
magnetostatic model of the magnetic network element.
\item
Non-WKB wave transport equations, incorporating linear reflection
terms, were solved for a grid of wave periods between 3 and
$3 \times 10^5$ s over the range of heights given above.
Below a mid-chromospheric {\em merging height} the waves were
modeled as modified kink-mode flux tube waves, and above this
height they were modeled as Alfv\'{e}n waves propagating parallel
to the background magnetic field.
\item
The waves are reflected strongly at the transition region, with
only about 5\% of the wave energy transmitted and 95\% reflected.
Above the transition region, most periods are reflected only
weakly by the large-scale radial gradient of the Alfv\'{e}n speed,
but periods exceeding 1 day are reflected strongly in
interplanetary space.
The period-averaged reflection coefficient from the extended
corona ($r = 2 \, R_{\odot}$) to 1 AU ranges between $10^{-4}$
and $10^{-3}$.
\item
At the photosphere, the wave periods containing the most power
are between 1 and 30 minutes, but the kink-mode waves with
periods greater than about 12 minutes are evanescent below the
merging height.
Only the shorter periods propagate to larger distances, and
above the merging height the power spectrum is dominated by
wave periods between about 1 and 6 minutes.
In~situ observations of power spectra that are strongest at
periods of several hours may be explained by many uncorrelated
flux tubes rotating past spacecraft in the solar wind on these
longer timescales.
\item
The period-averaged transverse velocity amplitude of the waves
agrees with observed off-limb nonthermal line widths from
SUMER and UVCS when the MBP jump amplitude $\sigma_j$ is taken
to be $\sim$3 km s$^{-1}$.
For all reasonable values of $\sigma_j$ between 0 and
6 km s$^{-1}$ the modeled wave amplitudes at heights greater
than 0.3 AU are significantly larger than the in~situ measurements,
implying that large-scale damping is needed.
\item
We investigated the potential impact of the linear viscous
damping of Alfv\'{e}n waves.
If the effective viscous stress timescale undergoes a
transition from its collisionally dominated form to either of
the possible collisionless forms (given by the $m=1$ or 2
cases of {\S}~6.2.1), the resulting damping lengths are
always much longer than the local height from the photosphere,
and viscosity thus cannot damp the waves.
\item
We also considered nonlinear turbulent damping using a
phenomenological model of the energy loss terms in the wave
transport equations.
The one free parameter is the normalization of the outer-scale
correlation length $L_{\perp}$ of the turbulence, which scales
with height as $B_{0}^{-1/2}$.
Interestingly, a single choice for the constant $\Lambda = 0.35$,
which specifies the value of $L_{\perp}$ at the merging height
to be about 1100 km, produces both the right amount of damping
above 0.3 AU to agree with the in~situ measurements and the
right amount of heating in the extended corona to agree with
empirically constrained solar wind acceleration models.
\item
Because of the relatively weak degree of reflection in the
corona, non-WKB effects are probably not important for computing
the wave pressure acceleration in regions that contribute to
the overall solar wind acceleration.
However, non-WKB effects produce order-of-magnitude differences
(from the WKB approximation) in the wave pressure acceleration
at and below the transition region.
The inclusion of damping also affects the total amount of
wave pressure acceleration imparted to the solar wind.
\end{enumerate}

\section{Discussion}

The model presented in this paper contains a great deal of
the relevant physics of Alfv\'{e}n waves in the solar atmosphere
and solar wind, but it is incomplete in several ways.
In this section we discuss various ways to
extend the model to improve its physical realism.

Our cylindrically symmetric magnetostatic model of the
network flux tube is highly idealized.
A fully three-dimensional model of a network ``patch,''
possibly using a high-resolution magnetogram image as the
lower boundary condition, would have a more realistic
geometry (see, e.g., van Ballegooijen \& Hasan 2003).
Such a model would contain a {\em distribution} of flux
tube strengths, merging heights, and canopy heights.
The resonant node structure seen in the power spectra
(e.g., Figure 8) would likely disappear in such a
heterogeneous model.

Our adopted photospheric boundary condition for transverse
flux-tube motions is also an idealized approximation.
First, the observations used to derive the velocity
autocorrelation functions are limited in time resolution.
Extremely high-frequency acoustic modes---e.g., with periods
{\em below} 1 minute---have been seen in the photosphere and
chromosphere (Deubner 1976;
Wunnenberg, Kneer, \& Hirzberger 2002;
DeForest, DePontieu, \& Hassler 2003),
and the kink-mode spectrum may extend to these frequencies as
well.
Second, much more needs to be learned about the statistics of
the ``jumps'' that we modeled as identical Gaussians.
A well-defined algorithm for measuring the combined power
spectrum of both isolated MBP motions and their mergings and
fragmentings should be developed.
It is possible that the transient nature of the motions may be
better represented by waves having both complex frequency and
complex wavenumber (e.g., Wang, Ulrich, \& Coroniti 1995), or
perhaps by suitably defined wavelet functions.
Also, new diagnostics such as polarization variability within
line profiles may add better constraints to our understanding
of flux-tube motions (Ploner \& Solanki 1997).

Compressible MHD waves and shocks must also be considered
alongside the incompressible waves modeled in this paper
(for recent work, see Rosenthal et al.\  2002;
Bogdan et al.\  2002, 2003; Hasan et al.\  2003;
Hasan \& Ulmschneider 2004; Wedemeyer et al.\  2004;
Bloomfield et al.\  2004; Suzuki 2004).
Longitudinal flux-tube waves are likely to be excited at or
below the photosphere, and these motions can provide additional
heating (via collisional damping) and, in some regions, they
can transfer some of their energy into {\em new} Alfv\'{e}nic
fluctuations that were not taken into account in our models.
In the solar wind, additional linear and nonlinear couplings
need to be included to understand better the relatively large
measured values of $\langle Z_{+} \rangle$ (see Figure 11).

The energy and momentum deposition terms computed in {\S}~6.3
need to be included in self-consistent solar wind models.
Extensive modeling has been done using isotropic Kolmogorov
turbulent heating terms (effectively, eq.~[\ref{eq:kolm}])
and the impact of a more realistic treatment of wave reflection
is likely to be significant.
Models including these effects can also be extended to the
slow solar wind associated with streamers, to coronal holes
at other phases of the solar cycle (e.g.,
Miralles, Cranmer, \& Kohl 2002), and possibly to
quasi-steady structures like spicules and prominences.

A substantial unresolved issue is how the wave-related heating
and acceleration is apportioned to electrons, protons, and
heavy ions in regions where these species become decoupled.
Our original motivation for this work was the determination
of ``outer-scale'' (i.e., low-frequency) Alfv\'{e}n wave
properties in the corona and solar wind, for use as initial
conditions in models of MHD turbulent cascade.
The kinetic consequences of this turbulence have been studied
recently by, e.g., Leamon et al.\  (1999, 2000),
Cranmer \& van Ballegooijen (2003),
Voitenko \& Goossens (2003, 2004), Gary \& Nishimura (2004),
and Gary \& Borovsky (2004).
The ultimate goal of this work is to model the spatial
evolution of a reflecting, cascading, and dissipating power
spectrum of fluctuations as a function of $k_{\parallel}$
and $k_{\perp}$.
Observations of preferential ion heating and acceleration,
as well as departures from isotropic Maxwellian distributions,
are key discriminators between competing models.

\acknowledgments

This paper is dedicated to the memory of Peter Nisenson, who
played a major role in the observations of G-band bright point
motions that made this work possible.
The authors would like to thank Eugene Avrett
for the use of tabulated PANDORA model atmospheres.
We are also grateful to
John Kohl, Leo Milano, William Matthaeus, Marco Velli,
Wolfgang Kalkofen, S. Peter Gary, and the anonymous referee
for valuable and inspiring discussions.
This work is supported by the National Aeronautics and Space
Administration (NASA) under grants NAG5-11913, NAG5-12865,
NAG5-10996, and NNG04GE84G to the Smithsonian Astrophysical
Observatory, by Agenzia Spaziale Italiana, and by the Swiss
contribution to the ESA PRODEX program.
This work has made extensive use of NASA's Astrophysics
Data System.

\begin{center}
{\bf APPENDICES}
\end{center}

\appendix

\section{Isothermal Kink-Mode Wave Properties}

In this Appendix we discuss a class of analytic solutions to
the incompressible thin-tube wave equations given in {\S}~5.1.
For constant coefficients in eq.~(\ref{eq:momSpruit}), we
assume that $v_{\perp}$ depends on time and height as
$e^{i \omega t - ikz}$.
For nonzero velocity amplitudes, this wave equation becomes
a simple quadratic equation in $k$ that is equivalent in
many ways to the acoustic-gravity wave equation
for an isothermal hydrostatic atmosphere (e.g., Lamb 1932;
Mihalas \& Mihalas 1984; Wang et al.\  1995).
For a real frequency $\omega$, this equation is satisfied
only for a complex $k$, with
\begin{equation}
  k \, = \, \frac{\omega_c}{V_{\rm ph}} \left( i \pm
  \sqrt{\frac{\omega^2}{\omega_{c}^2} - 1} \right)
  \label{eq:isoksol}
\end{equation}
where
\begin{equation}
  \omega_{c} \, \equiv \, \frac{g | \Delta \rho |}
  {2 V_{\rm ph} \, \rho_{\rm tot}}  \, = \,
  \sqrt{\frac{g}{8 H (2 \beta + 1)}}
  \label{eq:isoomc}
\end{equation}
is the critical kink-mode cutoff frequency.
The second expression above applies for an isolated
flux tube in an isothermal atmosphere with scale height
$H = a^{2}/g$, where $a$ is the isothermal sound speed
and $\beta$ is the ratio of gas pressure to magnetic pressure
(e.g,. Spruit 1981).
In this case, the identity $V_{\rm ph} / \omega_{c} = 4H$
yields the standard form of eq.~(\ref{eq:isoksol}).
Other useful identities in the isothermal limit, all
derivable from transverse pressure balance, are
\begin{equation}
  \frac{\rho_e}{\rho} = \frac{\beta + 1}{\beta}
  \,\, , \,\,\,\,\,
  \beta V_{A}^{2} = 2gH
  \,\, , \,\,\,\,\,
  B_{0} = \sqrt{8\pi a^{2} | \Delta \rho |}
  \,\, .
\end{equation}
Only for $\omega > \omega_{c}$ can there be a real component
of $k$, and thus a propagating wave.
A purely imaginary $k$ corresponds to evanescence.
 
For the background plasma state described in {\S}~3,
$\rho_{e} / \rho = 2.35$ and $V_{\rm ph} = 6.672$ km~s$^{-1}$
at the photosphere.
With these values, the critical period ($2\pi / \omega_c$)
is found to be 12.49 minutes.
Above the photosphere, the critical period computed with
eq.~(\ref{eq:isoomc}) decreases to a minimum
of 9.53 minutes at a height of $\sim$200 km, then begins to
increase, formally diverging to infinity at the merging height
where $\Delta \rho \rightarrow 0$ (see eq.~[\ref{eq:delrho}]).
This range of periods compares favorably with the chromospheric
estimates of 11.7 minutes given by Spruit (1981), derived for
$\beta = 1$, and 9 minutes given by Hasan \& Kalkofen (1999),
derived for $\beta = 0.3$.
For reference to other isothermal models, we use the above
relations to derive effective isothermal values of the plasma
beta and scale height at the photosphere; these are
$\beta = 0.74$ and $H=195$ km, although the latter is about
20\% larger than the actual density scale height at $z=0$.

The imaginary part of $k$ gives the time-averaged radial
dependence of the velocity amplitude.
For high frequencies corresponding to propagating kink-mode
waves, $| v_{\perp} |$ is proportional to $e^{z/4H}$.
(Because $\rho \propto e^{-z/H}$, this implies that
$| v_{\perp} | \propto \rho^{-1/4}$ as expected from WKB
Alfv\'{e}n wave theory.)
Thus, the kinetic energy density of the waves
integrated over the cross-section of the tube
(i.e., $\rho |v_{\perp}|^{2} / B_0$) is constant.\footnote{%
In the isothermal thin-tube limit, the vertical zero-order
magnetic pressure ($B_{0}^{2} / 8\pi$) must remain
proportional to the local gas pressure ($\rho a^2$),
so that $B_{0} \propto \exp (-z/2H)$.}
For low frequencies in the evanescent domain, let us define
$q \equiv (\omega / \omega_{c}) < 1$, and there are two
solutions for the radial dependence of the amplitude:
\begin{equation}
  | v_{\perp} | \, \propto \, \exp \left[ \frac{z}{4H}
  \left( 1 \pm \sqrt{1 - q^2} \right) \right]
  \,\,\, .
\end{equation}
The upper-sign solution is steeper than the propagating
solution, and the lower-sign solution is shallower.
In the limit $q \rightarrow 0$, the steep solution goes as
$\rho^{-1/2}$ and the shallow solution is constant.
Thus, for the steep [shallow] solution, the wave energy
density integrated over the tube area grows [decays] with
increasing height.
In {\S}~6 we see that non-WKB low-frequency waves naturally
``find'' the shallow solution, probably because it is the more
physical solution with decaying---not diverging---energy
density (see also Wang et al.\  1995).

An additional reason that a stellar atmosphere may ``choose''
the shallow evanescent solution instead of the steep solution
was given by Cranmer (1996) for acoustic-gravity waves in a
nonmagnetized atmosphere.
Despite the fact that the solar wind is extremely
{\em subsonic} in the deep photospheric and chromospheric
layers we study here, the existence of a nonzero gradient
of the outflow speed leads to additional terms in the wave
dispersion relation.
Cranmer (1996) showed that these terms yield a nonzero
real part of $k$ for all frequencies, even those below the
evanescent cutoff.
For a subsonic flow ($\epsilon \equiv u/a \ll 1$) in
the evanescent regime ($q \ll 1$) the real part of the radial
wavenumber is given by
\begin{equation}
 k_{r} \, \approx \, \left\{
  \begin{array}{rl}
    -3 \epsilon q / H  \,\,\, , & \mbox{steep} \\
    +  \epsilon q / H  \,\,\, , & \mbox{shallow}
  \end{array}
 \right.
\end{equation}
and the shallow solution is the one with a positive (i.e.,
upward) phase speed $\omega / k_r$ corresponding to the
physically relevant situation of more upward then downward
wave power.

Lastly we compute the frequency-dependent partition between
kinetic energy density ($E_{K} = \rho |v_{\perp}|^{2}/2$)
and magnetic energy density ($E_{B} = |B_{\perp}|^{2}/ 8\pi$).
The linearized induction equation (eq.~[\ref{eq:indSpruit}])
is written in the isothermal limit as
\begin{equation}
  \omega B_{\perp} \, = \, -k B_{0} v_{\perp}
\end{equation}
and thus the ratio of energies is
\begin{equation}
  \frac{E_B}{E_K} \, = \,
  \frac{V_{\rm ph}^{2} | k |^2}{\omega^2} \,\,\, .
\end{equation}
Using the analytic solution for $k$ above, the total energy
density is given by
\begin{equation}
  E_{\rm tot} (\omega) \, = \, E_{K} (\omega) \times \left\{
  \begin{array}{ll}
    2 &
    , \,\,\, q \geq 1 \\
    1 + \left( 1 \pm \sqrt{1 - q^{2}} \right)^{2} / q^{2} &
    , \,\,\, q < 1
  \end{array} \right.
  \label{eq:isopart}
\end{equation}
where the upper [lower] sign gives the steep [shallow]
evanescent solution as defined above.
Note that in the limit $q \rightarrow 0$, the shallow solution
is dominated by kinetic fluctuations and the steep solution is
dominated by magnetic fluctuations.

\section{Non-WKB Alfv\'{e}n Wave Transport Equations}

Over the years there have been several versions of the the
non-WKB Alfv\'{e}n wave equations published in the literature
that superficially do not resemble one another.
This introduces a potential difficulty in comparing the results
of these past efforts.
In this Appendix we collect several of these differently
appearing equations and confirm that they are essentially
equivalent.

The transport equations of Heinemann \& Olbert (1980) are
expressible in terms of our Elsasser variables
(eq.~[\ref{eq:elsdefine}]) as
\begin{equation}
  \left[ \frac{\partial}{\partial t} + ( {\bf u} \mp
  {\bf V}_{A} ) \cdot \nabla \right] z_{\pm} \, = \,
  z_{\pm} \left[ ( {\bf u} \pm {\bf V}_{A} ) \cdot \nabla \right]
  \ln \left( \rho^{1/4} \right) -
  z_{\mp} \left[ ( {\bf u} \pm {\bf V}_{A} ) \cdot \nabla \right]
  \ln \left( \rho^{1/4} B_{0}^{-1/2} \right)
\end{equation}
where we have specified that the transverse length scale of the
flux tube is proportional to $B_{0}^{-1/2}$.
It is relatively easy to see the correspondence between the
above form and our eq.~(\ref{eq:zpm}) when we assume that the
vectors ${\bf u}$ and ${\bf V}_A$ are purely radial and that
${\bf z}_{\pm}$ is perpendicular to the radial direction.
For mass conservation in a flux tube, the quantity
$\rho^{1/4} B_{0}^{-1/2}$ is proportional to $V_{A}^{-1/2}$,
and thus the above logarithmic derivatives yield the scale heights
defined in {\S}~5.2.
Khabibrakhmanov \& Summers (1997) gave essentially the same
equations, but allowing for more general wave polarization states
and pressure anisotropy.

The Alfv\'{e}nic transport equations of Zhou \& Matthaeus (1990)
contain nonlinear terms, and thus in some ways are more general
than the equations discussed in this paper.
However, when looking only at their linear terms, one obtains
\begin{equation}
  \frac{\partial {\bf z}_{\pm}}{\partial t} +
  ( {\bf u} \mp {\bf V}_{A} ) \cdot \nabla {\bf z}_{\pm}
  \, = \,
  \left( \frac{{\bf z}_{\mp} - {\bf z}_{\pm}}{2} \right)
  \nabla \cdot \left( \frac{\bf u}{2} \pm {\bf V}_{A}
  \right) - {\bf z}_{\mp} \cdot \left(
  \nabla {\bf u} \pm \frac{\nabla {\bf B}_0}{\sqrt{4\pi\rho}}
  \right)
  \label{eq:zhoum90}
\end{equation}
(see also Dmitruk, Milano, \& Matthaeus 2001).
The above equation is exactly equivalent to eq.~(\ref{eq:zpm}),
but some algebraic manipulation is required in order to
demonstrate this.
Among other steps, one needs to apply the auxiliary conditions
given by Zhou \& Matthaeus (1990),
\begin{equation}
  {\bf u} \cdot \nabla \left( \frac{1}{\sqrt{4\pi\rho}}
  \right) \, = \, \frac{1}{\sqrt{4\pi\rho}}
  \nabla \cdot \left( \frac{\bf u}{2} \right)
  \,\,\, ,
\end{equation}
together with the fact that $\nabla \cdot {\bf B}_{0} = 0$.

The non-WKB equations given by Velli (1993) are similar
to those of Zhou \& Matthaeus (1990).
Recasting Velli's (1993) differently-defined Elsasser variables
into our conventional form, one obtains
\begin{equation}
  \frac{\partial {\bf z}_{\pm}}{\partial t} +
  ( {\bf u} \mp {\bf V}_{A} ) \cdot \nabla {\bf z}_{\pm}
  \, = \,
  \left( \frac{{\bf z}_{\mp} - {\bf z}_{\pm}}{2} \right)
  \nabla \cdot \left( \frac{\bf u}{2} \pm {\bf V}_{A}
  \right) - {\bf z}_{\mp} \cdot \nabla \left(
  {\bf u} \pm {\bf V}_{A} \right)
  \,\,\, .
\end{equation}
Only the last term differs from eq.~(\ref{eq:zhoum90}) above,
but---at least for the incompressible transverse waves we
study here---they are equivalent.
This can be shown by noting that, for an arbitrary scalar
function $\psi (r)$, one can state that
\begin{equation}
  {\bf z} \cdot \nabla (\psi {\bf B}) \, = \,
  \psi \, {\bf z} \cdot \nabla {\bf B}
\end{equation}
only when two conditions are met: (1) the field is divergence-free
($\nabla \cdot {\bf B} = 0$), and (2) the fluctuations are
purely transverse (i.e., when ${\bf z} \cdot \nabla \psi = 0$).
In the above comparison, this condition is satisfied because
$\psi = 1 / \sqrt{4\pi\rho}$ depends only
on the radius $r$ and the incompressible ${\bf z}$ amplitudes
have a zero $r$-component.

Barkhudarov (1991) derived the non-WKB transport equations in
terms of nonstandard Elsasser-like variables
$I_{\pm} \equiv ( B_{\perp} \pm v_{\perp}\sqrt{4\pi\rho} )$.
Also, in that paper the equations were given only for spherical
symmetry.
For a more general superradially expanding flux tube, the
equations of Barkhudarov (1991) become
\begin{equation}
  \frac{\partial I_{\pm}}{\partial t} +
  (u \mp V_{A})
  \frac{\partial I_{\pm}}{\partial r} \, = \,
  -I_{\pm} \left( \frac{\partial u}{\partial r} -
  \frac{u \mp V_{A}}{2 B_{\rm o}}
  \frac{\partial B_{\rm 0}}{\partial r}
  \right)
  \mp \frac{u \pm V_{A}}{2 V_A}
  \frac{\partial V_A}{\partial r}  (I_{+} - I_{-})
\end{equation}
and these are the specific equations solved by the numerical
Runge-Kutta code described in {\S}~5.2.

For completeness, we also give the full equation of
{\em wave action conservation} (i.e., wave energy conservation
measured in the comoving wind frame).
We use the variable naming convention of Barkhudarov (1991),
where $f$ and $g$ correspond to outward and inward
waves, respectively, but we follow Heinemann \& Olbert (1980)
in defining these quantities in velocity units and for general
superradial expansion:
\begin{equation}
  f^{2} \, \equiv \, \frac{|z_{-}|^{2} (u + V_{A})^2}{u V_A}
  \,\,\, , \,\,\,\,\,\,\,
  g^{2} \, \equiv \, \frac{|z_{+}|^{2} (u - V_{A})^2}{u V_A}
  \,\,\, .
\end{equation}
The wave action conservation equation is obtained by writing
the four real transport equations (for the real and imaginary
parts of $z_{+}$ and $z_{-}$) and multiplying each term in
the equations by their corresponding real or imaginary
Elsasser component.
The sum of all four equations yields the wave action equation:
\begin{equation}
  \frac{\partial}{\partial t} \left[ \rho u \left(
  \frac{f^2}{u + V_A} - \frac{g^2}{u - V_A} \right) \right]
  + \nabla \cdot \left[ \rho {\bf u} ( f^{2} - g^{2} ) \right]
  \, = \, 0  \,\,\, .
\end{equation}
In regions dominated by outward propagating waves
($|z_{-}| \gg |z_{+}|$), where both the background properties
and wave amplitudes are time-steady, this reduces to simply
\begin{equation}
  \frac{\partial f^2}{\partial r} \, = \, 0
  \label{eq:wvact}
\end{equation}
and the assumption of wave action conservation is equivalent
to the assumption that $f$ is constant
(see also Jacques 1977; Barkhudarov 1991).
This is not necessarily a WKB result, but it applies only in
regions of negligible reflection.

\section{Issues Relating to Anisotropic MHD Turbulence}

In the wave transport models presented in this paper we have
not been overly concerned with the direction of the Alfv\'{e}n
wavevector ${\bf k}$ relative to the background magnetic
field ${\bf B}_0$.
Mainly this is because the dispersion relation of the waves
in the MHD limit depends only on $k_{\parallel}$, the parallel
component of the wavevector,
\begin{equation}
  \omega \, = \, (u \mp V_{A}) k_{\parallel}
\end{equation}
for the $z_{\pm}$ modes, and not on the component $k_{\perp}$
perpendicular to the field.
Thus the monochromatic wave transport equations ``know''
about $k_{\parallel}$ via their dependence on wave frequency
$\omega$ but are independent of $k_{\perp}$.
In this Appendix we investigate several issues concerning
the expected distribution of propagation directions and its
evolution as MHD turbulent cascade develops.

From a linear standpoint, an initial distribution of
propagation directions $\theta$ (i.e., the angle between
${\bf k}$ and ${\bf B}_0$) will evolve with heliocentric
distance due to several physical processes.
The radial variations of the inertial-frame phase speeds
$u \mp V_A$ induce a radial ``stretching'' in wavelength and
thus a large-scale decrease in $k_{\parallel}$ with increasing
height.
The transverse spreading of the open flux tube should also
cause the perpendicular wavenumber $k_{\perp}$ to decrease with
increasing height as well.
The overall sense of increase or decrease in $\theta$ depends
on the time-steady properties of the magnetic field and outflow.
Other effects can contribute to a radial variation in $\theta$.
First, any transverse gradients of the Alfv\'{e}n speed can
refract oblique waves toward the center of the flux tube,
thus decreasing $\theta$ (Wentzel 1989).
Second, viscous damping of high-frequency waves preferentially
dissipates oblique fluctuations, since
$L_{c} \propto \cos^{2} \theta$.
The analysis in {\S}~6.2.1 assumed $\theta = 0$, but the
impact of viscous damping can be enhanced significantly if
the waves are sufficiently oblique.
All of the above effects should be included if the radial
variation of $\theta$ (or $\langle \theta \rangle$ averaged
over the power spectrum) is to be computed.

As the turbulent cascade develops in the corona, an additional
nonlinear evolution of the wave energy within ``wavenumber
space'' ($k_{\parallel}, k_{\perp}$) occurs.
It has been known for several decades that MHD turbulence
in the presence of a steady-state magnetic field develops
a strong anisotropy in wavenumber.
Both numerical simulations and RMHD analytic descriptions
indicate that the spectral transport from large to small
spatial scales proceeds mainly in the two-dimensional plane
perpendicular to ${\bf B}_0$ (see, e.g., Higdon 1984;
Shebalin, Matthaeus, \& Montgomery 1983;
Goldreich \& Sridhar 1995, 1997; Bhattacharjee \& Ng 2001;
Cho, Lazarian, \& Vishniac 2002; Oughton et al.\  2004).
That is, the cascade spreads out the power to successively
larger values of $k_{\perp}$ while leaving $k_{\parallel}$
relatively unchanged.

Here we follow Cranmer \& van Ballegooijen (2003) and give
an analytic solution for the turbulent power spectrum in the
limit that the cascade is allowed to proceed to its final
``driven'' steady state in a small homogeneous volume of
plasma.
We then compare the frequency dependence of this power spectrum
to the empirically constrained (linear) frequency spectrum
as presented in Figure 8.
Let us define the three-dimensional total power spectrum as
\begin{equation}
  U_{\rm tot} \, = \,
  \int d^{3} {\bf k} \, P_{\rm 3D} ({\bf k})
  \label{eq:P3Dint}
\end{equation}
where we write the volume element $d^{3} {\bf k}$ in cylindrical
coordinates as $2\pi k_{\perp} \, dk_{\perp} dk_{\parallel}$
and thus assume symmetry in the two directions transverse to
the background field.
Cranmer \& Ballegooijen (2003), following the general development
of RMHD anisotropic cascade theory, assumed that the turbulence
eventually leads to a state of complete mixing between the inward
and outward modes, and thus $U_{+} = U_{-} = U_{\rm tot}/2$.
This assumption is incompatible with the non-WKB wave reflection
models computed in this paper, but we follow the earlier analysis
in order to compare the {\em shapes} of the relevant power spectra
and to obtain insight about how to better model the turbulent
cascade.

For fully developed anisotropic MHD turbulence, we assume the
power spectrum to be a separable function of two variables:
$k_{\perp}$ and a nonlinearity parameter $y$ defined as the
ratio of the local wind-frame frequency $V_{A} k_{\parallel}$ to
an assumed nonlinear eddy turnover rate
$\langle \delta V \rangle k_{\perp}$ (see, e.g.,
Goldreich \& Sridhar 1995).
We use the notation of {\S}~2.3 of
Cranmer \& van Ballegooijen (2003) and define
\begin{equation}
  P_{\rm 3D} (k_{\parallel}, k_{\perp}) \, = \,
  \frac{\rho V_{A} W_{\perp}^{1/2}}{k_{\perp}^3} \, g(y)
  \label{eq:P3D}
\end{equation}
where $W_{\perp} (k_{\perp})$ is a reduced power spectrum,
scaled to velocity-squared units and defined as
\begin{equation}
  W_{\perp} (k_{\perp}) \, = \, \frac{k_{\perp}^2}{\rho}
  \int_{-\infty}^{+\infty} dk_{\parallel} \,
  P_{\rm 3D} (k_{\parallel}, k_{\perp})  \,\,\, .
\end{equation}
For the MHD inertial range, $W_{\perp}$ is proportional
to $k_{\perp}^{-2/3}$.
For simplicity, let us define
\begin{equation}
  W_{\perp} (k_{\perp}) \, = \, \left\{
  \begin{array}{ll}
    0  \,\, , & k_{\perp} < k_{\rm out} \\
    U_{\rm tot} ( k_{\perp} / k_{\rm out} )^{-2/3}
      (3\pi\rho)^{-1} \,\, ,
      & k_{\perp} \geq k_{\rm out}
  \end{array} \right.
  \label{eq:wperpout}
\end{equation}
for a given outer-scale perpendicular wavenumber $k_{\rm out}$.
The factor of $3 \pi \rho$ above is needed to normalize the
full power spectrum as defined in eq.~(\ref{eq:P3Dint}).

The $k_{\parallel}$ dependence of the power spectrum is
contained in the dimensionless $g(y)$ function in
eq.~(\ref{eq:P3D}).
The condition $y=1$ is defined as ``critical balance'' by
Goldreich \& Sridhar (1995), and their analysis only
constrains the general form of $g(y)$, not its exact value.
The bulk of the wave power is believed to reside at low values
of $y$ (e.g., $g$ is nonzero only for $|y| \lesssim 1$),
where we define
\begin{equation}
  y \, = \, \frac{k_{\parallel} V_A}{k_{\perp} W_{\perp}^{1/2}}
  \,\,\, .
\end{equation}
This condition captures the highly nonlinear state of
turbulence, for which a coherent wave survives for no more
than about one period before nonlinear processes transfer its
energy to smaller scales.
One reasonable possibility for the parametric dependence of
this function is a normalized Gaussian,
\begin{equation}
  g(y) \, \approx \, \frac{1}{\sqrt{\pi}} \,
  e^{-y^2}  \,\,\, .
  \label{eq:ggauss}
\end{equation}
Cho et al.\  (2002) analyzed numerical simulations of MHD
turbulence and found that $g(y)$ can be fit reasonably well
by decaying exponential or Castaing functions.
Cranmer \& van Ballegooijen (2003) solved a simple wavenumber
diffusion equation to obtain the analytic form
\begin{equation}
  g(y) \, = \, \frac{\Gamma (n)}{\Gamma(n - 0.5)}
  \sqrt{\frac{\gamma ( 1 - \zeta )^2}{\pi \alpha}}
  \left[ 1 + \frac{\gamma ( 1 - \zeta )^{2} y^{2}}{\alpha}
  \right]^{-n}
  \label{eq:gkappa}
\end{equation}
which is normalized to unity when integrated over all $y$, and with
\begin{equation}
  n \, = \, \frac{(\beta / \gamma) + \zeta + 1}
  {2 (1 - \zeta)}
  \,\,\, .
\end{equation}
For the MHD inertial range, $\zeta = 1/3$, and the three constants
$\alpha$, $\beta$, and $\gamma$ describe the relative strengths
of $k_{\parallel}$ diffusion, $k_{\perp}$ advection, and
$k_{\perp}$ diffusion, respectively.
We assume $\alpha = \gamma$, thus leaving the only ``free''
parameter to be the ratio $\beta / \gamma$.
Cranmer \& van Ballegooijen (2003) discussed the most realistic
values of this ratio; earlier models implied that
$\beta / \gamma \approx 1$, but one would need this ratio
to be smaller than about 0.25 in order to produce enough
parallel cascade in the corona to heat protons and heavy ions
via cyclotron resonance.

With the above definitions it becomes possible to derive the
effective frequency power spectra $\widetilde{P}_{\pm}(\omega)$
that are consistent with fully developed anisotropic turbulence.
Noting that we define $\omega$ as the frequency measured in
the inertial frame centered on the Sun, we define
\begin{equation}
  \widetilde{P}_{\pm} (\omega) \, = \, \frac{1}{2}
  \int d^{3} {\bf k} \, P_{\rm 3D} ({\bf k}) \,\,
  \delta \left[ \omega - (u \mp V_{A}) k_{\parallel} \right]
\end{equation}
where the Dirac delta function collapses the three-dimensional
wavenumber spectrum into a one-dimensional frequency spectrum.
This ``translation'' must be done differently for the inward and
outward mode spectra, as indicated by the difference between the
inertial-frame phase speeds of the two modes.
The above expression is defined formally for both positive and
negative frequencies; below we multiply by 2 in order to
consider only positive frequencies.
For the purposes of this Appendix, let us use the simpler form
of eq.~(\ref{eq:ggauss}) for $g(y)$ that gives the
straightforward analytic solution
\begin{equation}
  \widetilde{P}_{\pm} (\omega) \, = \, \frac{U_{\rm tot} \,
  \omega_{{\rm out,} \pm}}{2 \pi^{1/2} \omega^2}
  \left\{ 1 - \exp \left[ - \left(
  \frac{\omega}{\omega_{{\rm out,} \pm}}
  \right)^{2} \right] \right\}
\end{equation}
where we define the effective outer-scale frequency
\begin{equation}
  \omega_{{\rm out,} \pm} \, = \,
  \frac{| u \mp V_{A} | k_{\rm out}}{V_A}
  \sqrt{\frac{U_{\rm tot}}{3\pi\rho}}
\end{equation}
and we assume $k_{\rm out} = 2\pi/L_{\perp}$.
We note that the use of eq.~(\ref{eq:gkappa}) for $g(y)$ would
have yielded a solution similar to the above form of
$\widetilde{P}_{\pm}$, only differing slightly in a small region
of frequency near $\omega_{{\rm out,} \pm}$.
This frequency sets the scale for the spectrum.
We use the effective outer-scale frequency for
outward propagating waves to define the  nonlinear ``driving time''
$t_{0} = 2\pi / \omega_{{\rm out,} -}$ plotted in Figure 14.

Figure 19 compares the above turbulent frequency power spectra
$\widetilde{P}_{\pm} (\omega)$ with the empirically constrained
linear frequency spectra $P_{\pm} (\omega)$ as derived
in {\S}~5 above.
For simplicity we plot the spectra at only one heliocentric
distance ($r = 2 \, R_{\odot}$) and note that one can use the
radial dependences of the various timescales plotted in Figure 14
in order to extend these results to other heights.
The above assumption of full mixing for the nonlinear cascade
(i.e., $U_{-} = U_{+}$) is far from the linear non-WKB result
of $U_{-} \gg U_{+}$ in the extended corona.
The inability of the analytic RMHD anisotropic cascade theory
to take account of the strong imbalance between inward and outward
wave amplitudes is a major shortcoming of this kind of model.
We need a better understanding of the {\em detailed wavenumber
dependence} of the inward-to-outward power ratio for MHD turbulence.

In Figure 19 we use the preferred normalization $\Lambda = 0.35$
to set the scale of $L_{\perp}$.
With this normalization at $r = 2 \, R_{\odot}$, the outer-scale
driving timescales $2\pi / \omega_{{\rm out,} \pm}$
are 5.7 and 4.1 minutes for the $+$ and $-$ signs, respectively.
The peaks of the scaled power spectra
$\omega \widetilde{P}_{\pm}$ occur at frequencies of
$\sim 1.12 \omega_{{\rm out,} \pm}$, but the empirically constrained
linear spectra $\omega P_{\pm}$ contain most of their power at
slightly higher frequencies (i.e., shorter periods).
The fact that the two spectra exhibit maxima within the same
order of magnitude of frequency, though, may be just a result
of choosing the radius of $r = 2 \, R_{\odot}$.
Figure 14 shows that if the curves in Figure 19 were computed
instead at 1 AU, the peaks of the linear $P_{\pm}$ curves
would be in the same range of periods (1 to 5 minutes), but the
peaks of the $\widetilde{P}_{\pm}$ curves would be at periods
between 30 and 40 minutes.
At 1 AU the bulk of the power in the linear frequency
spectrum would be at an extremely high local frequency
from the standpoint of the cascade (i.e., $y \gg 1$).
This suggests that the Sun emits a ``fossil'' frequency spectrum
that eventually can advect into a region of the three-dimensional
wavenumber spectrum that was previously believed to be strongly
{\em depleted} due to the anisotropic cascade.
The implications of this result will be investigated in
future work.

\clearpage
 
\begin{figure}
\epsscale{0.97}
\plotone{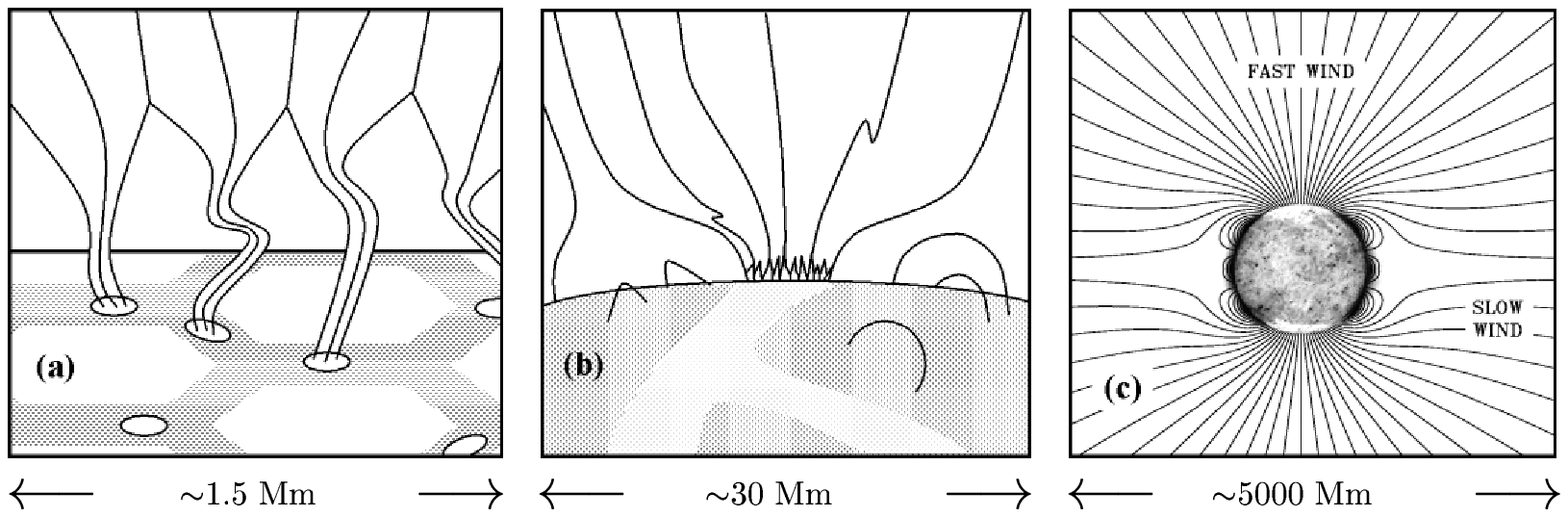}
\caption{\small
Summary sketch of the magnetic field structure discussed
in this paper, with the fields of view successively widening
from (a) to (b) to (c).
(a) Intergranular lanes host G-band bright points that are
shaken transversely to generate kink-mode waves.
(b) Above the height where individual flux tubes
merge, the coronal-hole network field is mainly open
(with a funnel/canopy structure), and kink-mode waves
are transformed into Alfv\'{e}n waves.
(c) Non-WKB waves in the solar wind propagate and reflect
depending on their frequencies, and MHD turbulent cascade
can occur where outward and inward waves interact nonlinearly.
The inverted solar image was obtained by EIT
(Extreme-ultraviolet Imaging Telescope) on {\em{SOHO}}
(e.g., Moses et al.\  1997).
Field lines in (c) are plotted at {2.1\arcdeg} intervals at
the solar surface, and thus each pair of lines encompasses
$\sim$1--2 network funnels.}
\end{figure}

\begin{figure}
\epsscale{0.49}
\plotone{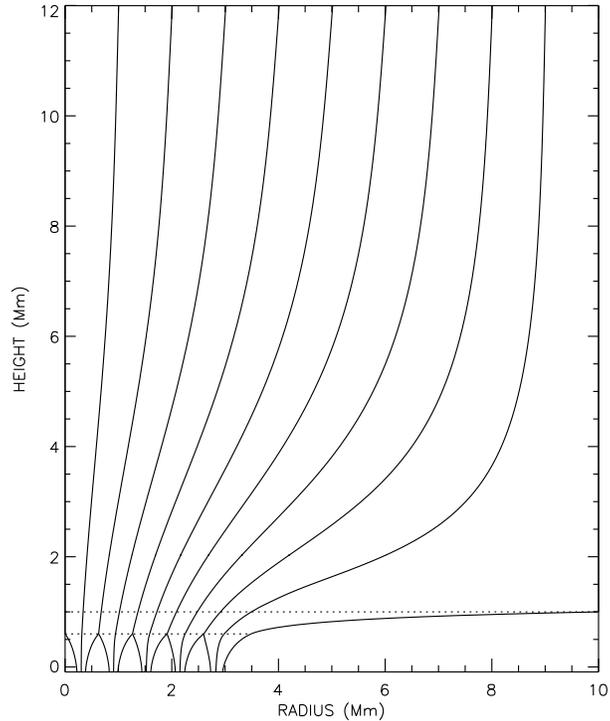}
\caption{\small
Magnetic field lines in a cross section of the magnetostatic
model of a network element.
Dotted lines at $z = 0.6$ and 1 Mm denote the merging height
$z_m$ and the canopy height $z_c$.
Above $z_m$ the model has cylindrical symmetry about the
left axis of the plot; below $z_m$ we model the flux tubes
as evenly distributed throughout the circular network patch
of radius $\sim$3 Mm, and the plotted cross sections are
shown only for illustrative purposes.}
\end{figure}

\clearpage
 
\begin{figure}
\epsscale{0.51}
\plotone{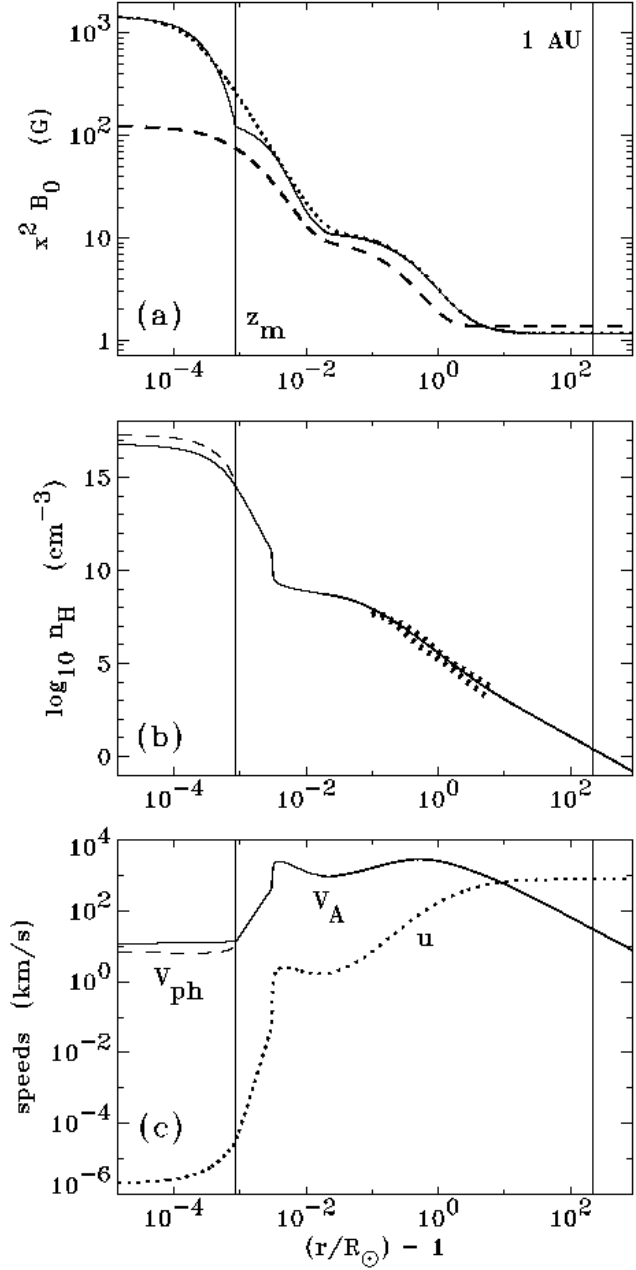}
\caption{\small
Steady-state plasma conditions along the central axis of
the network flux tube.
(a) Magnetic field strength, multiplied by $x^2$, for the
adopted model ({\em{solid line}}), the model of Hackenberg
et al.\  (2000) ({\em{dotted line}}), and the model of
Li (2003) ({\em{dashed line}}).
(b) Adopted model of the hydrogen number density
({\em{solid line}}), empirical limits taken from the minima
and maxima of observations by Guhathakurta \& Holzer (1994),
Fisher \& Guhathakurta (1995), and Doyle et al.\  (1999)
({\em{dotted lines}}), and the quantity
$\rho_{\rm tot}/1.2/m_{\rm H}$ ({\em{dashed line}}).
(c) Alfv\'{e}n speed $V_{A}$ ({\em{solid line}}),
outflow speed $u$ ({\em{dotted line}}), and flux tube
phase speed $V_{\rm ph}$ ({\em{dashed line}}).}
\end{figure}

\clearpage
 
\begin{figure}
\epsscale{1.00}
\plotone{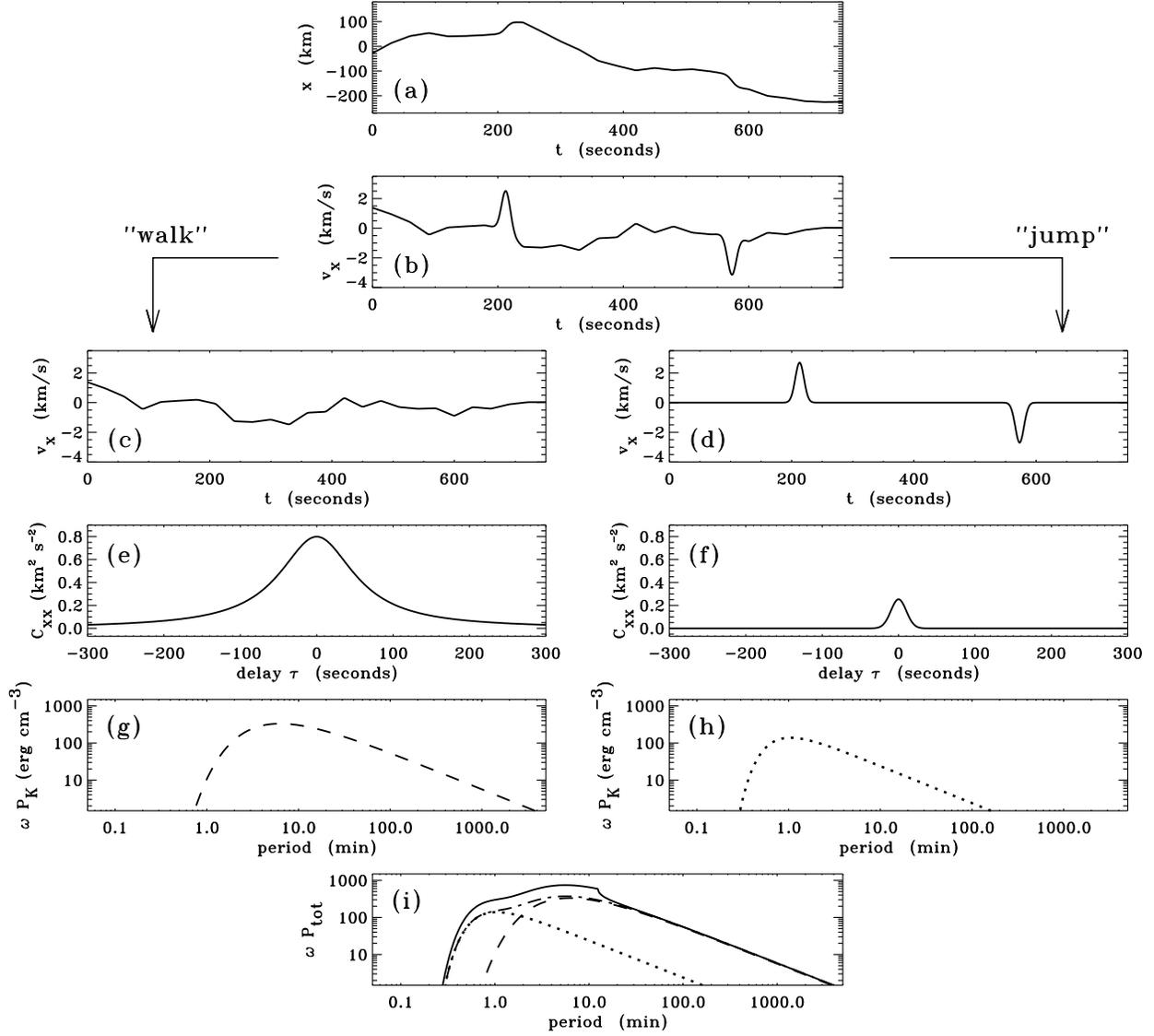}
\caption{\small
Outline of the empirical procedure used to derive the
photospheric power spectrum of MBP kink-mode wave energy.
Horizontal MBP positions (a) are differentiated to obtain
velocities (b), which are separated into ``walk'' and
``jump'' phases (c,d).
The autocorrelation functions (e,f) of these velocity time
series are computed and Fourier transformed to obtain
power spectra (g,h) as a function of frequency $\omega$,
here plotted versus period ($2\pi / \omega$) in minutes.
The kinetic energy power spectra from the statistically
independent walk ({\em{dashed lines}}) and
jump ({\em{dotted lines}}) phases are summed (i)
and used to compute the total energy spectrum
({\em{solid line}}).}
\end{figure}

\clearpage
 
\begin{figure}
\epsscale{0.5}
\plotone{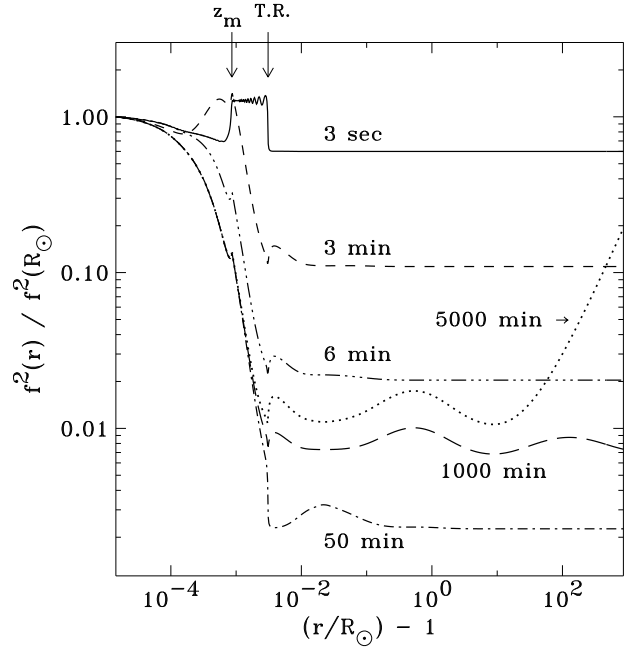}
\caption{\small
Height dependence of the outward propagating wave action flux
for a selection of periods, each normalized to its value in
the photosphere.  (See labels for values of the wave period.)
The merging height $z_m$ and the transition region (T.R.) are
labeled with arrows.}
\end{figure}

\begin{figure}
\epsscale{0.97}
\plotone{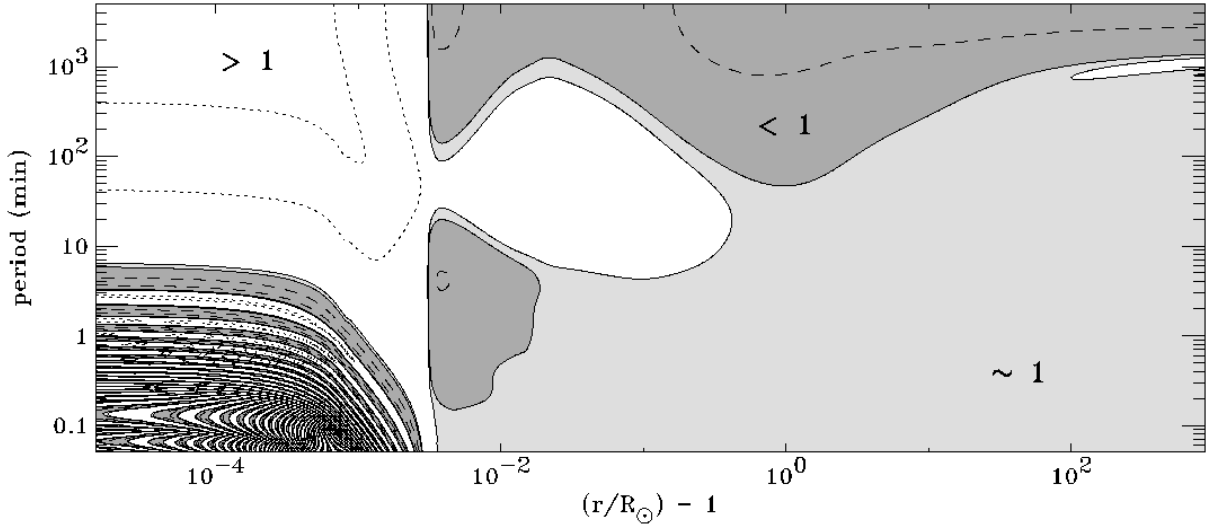}
\caption{\small
Contour/grayscale plot of the ratio $E_{K}/E_{B}$ as a
function of height and wave period.
The MHD-like regime of energy equipartition (i.e., a
ratio of nearly 1) is shown in light gray, bounded by solid
contours at values of 0.8 and 1.2.
The kinetic-dominated regime is in white with dotted
contours at values of $10^2$ and $10^4$.
The magnetic-dominated regime is in dark gray with dashed
contours at a value of 0.1.}
\end{figure}

\clearpage

\begin{figure}
\epsscale{0.97}
\plotone{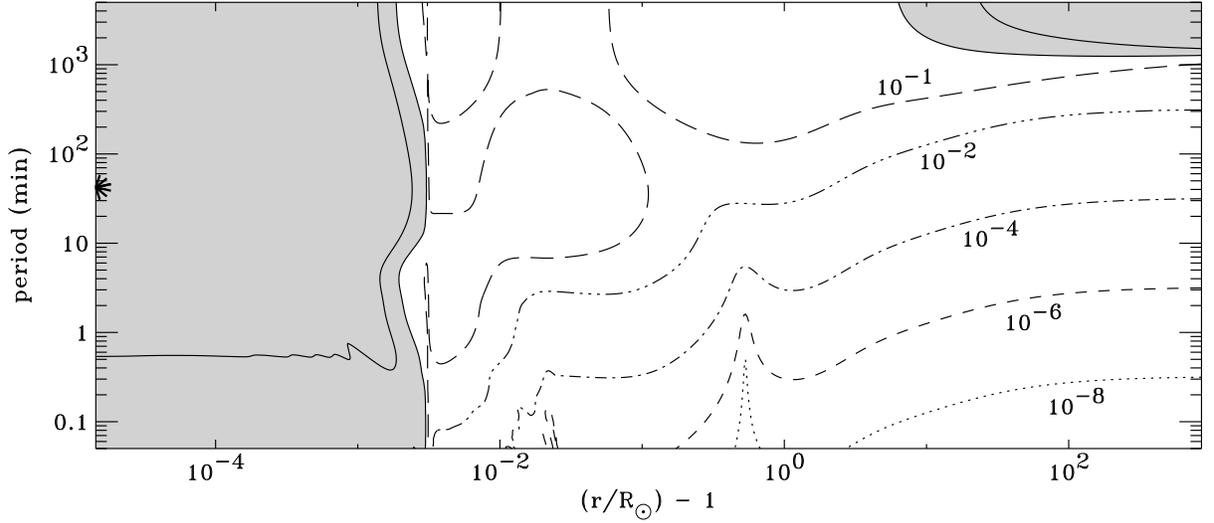}
\caption{\small
Contour/grayscale plot of the ratio $E_{+}/E_{-}$ as a
function of height and wave period.
The region of ``strong'' reflection is in gray, bounded
by solid contours at values of 0.5 and 0.75.
The location of the maximum value (0.99825) is denoted by
a star.  Other contours range between values of 10$^{-8}$
and 10$^{-1}$ and are labeled by their values.
The transition region is clearly seen
at $z \approx 0.003 \, R_{\odot}$.}
\end{figure}

\begin{figure}
\epsscale{0.5}
\plotone{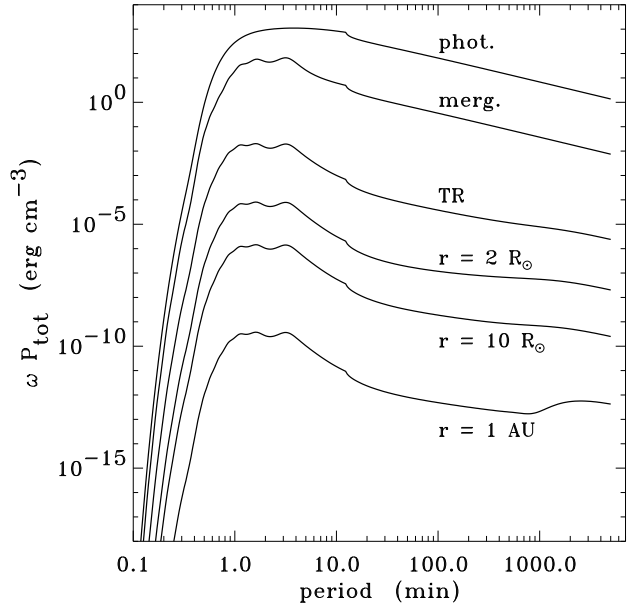}
\caption{\small
Period and height dependence of the scaled total power
spectrum $\omega P_{\rm tot} (\omega, r)$ for six
heights (top to bottom): $z = 0$, 0.6 Mm (merging
height), 2.15 Mm (transition region), 1 $R_{\odot}$,
9 $R_{\odot}$, and 214 $R_{\odot}$.
The latter 3 heights are labeled with their heliocentric
radii.  The MBP jump amplitude was assumed to be
$\sigma_{j} = 3$ km~s$^{-1}$.}
\end{figure}

\clearpage

\begin{figure}
\epsscale{1.00}
\plotone{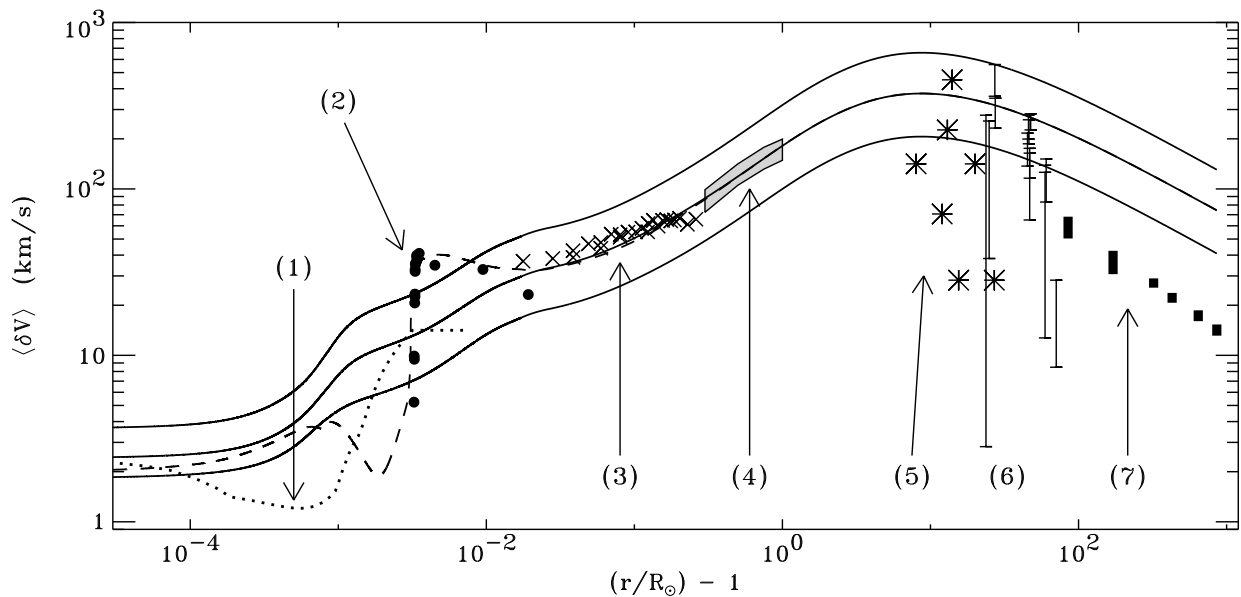}
\caption{\small
Height dependence of the frequency-integrated velocity
amplitude.  Solid lines give the undamped value of
$\langle \delta V \rangle$ for three choices of $\sigma_{j}$:
0, 3, and 6 km s$^{-1}$ (from bottom to top).
The dashed line is $\langle \delta V \rangle_{B}$ for
the $\sigma_{j} = 3$ km~s$^{-1}$ case.
Other lines and symbols correspond to observations
discussed in detail in {\S}~6.1.}
\end{figure}

\begin{figure}
\epsscale{0.5}
\plotone{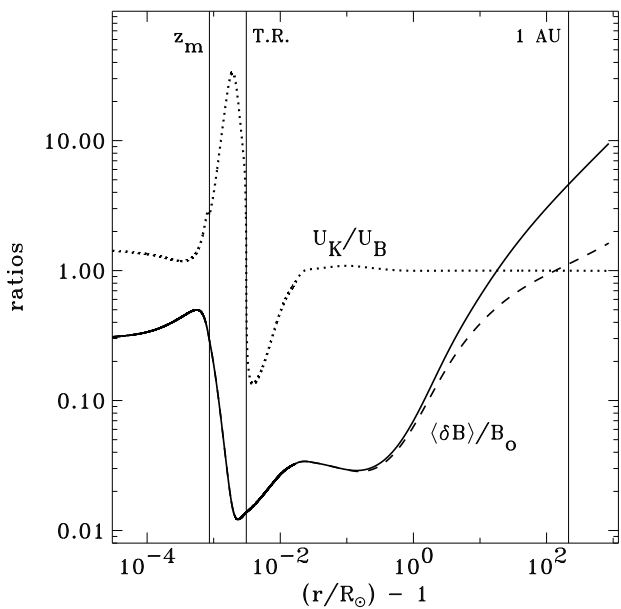}
\caption{\small
Frequency-integrated ratios for the baseline $\sigma_{j} = 3$
km~s$^{-1}$ model:
kinetic-to-magnetic energy density ratio ({\em{dotted line}}),
dimensionless magnetic amplitude for the undamped linear
model ({\em{solid line}}) and for the nonlinearly damped
$\Lambda = 0.35$ model ({\em{dashed line}}).
The merging height, transition region and 1 AU are labeled with
thin vertical lines.}
\end{figure}

\clearpage

\begin{figure}
\epsscale{0.47}
\plotone{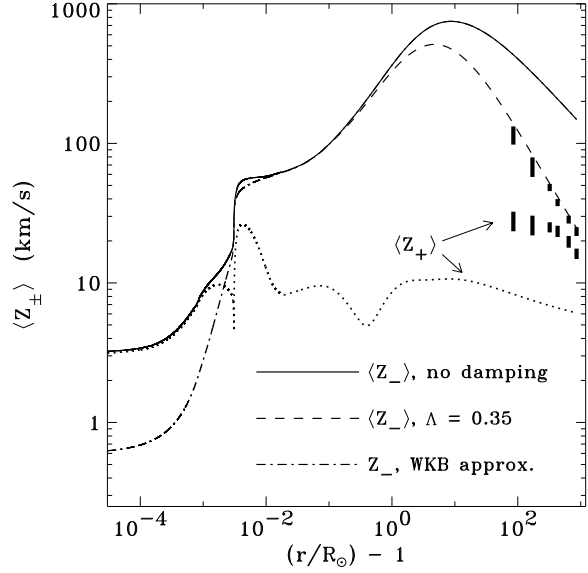}
\caption{\small
Height dependence of the frequency-integrated Elsasser
variables for the $\sigma_{j} = 3$ km~s$^{-1}$ model:
linear undamped $\langle Z_{-} \rangle$ ({\em{solid line}}),
nonlinearly damped $\langle Z_{-} \rangle$ ({\em{dashed line}}),
WKB height dependence of $Z_{-}$ ({\em{dot-dashed line}}),
and the linear value of $\langle Z_{+} \rangle$
({\em{dotted line}}).
In situ measurements from {\em Helios} and {\em Ulysses} are
shown as solid bars (Bavassano et al.\  2000), with the
upper set corresponding to $Z_{-}$ and the lower set to
$Z_{+}$.}
\end{figure}

\begin{figure}
\epsscale{0.47}
\plotone{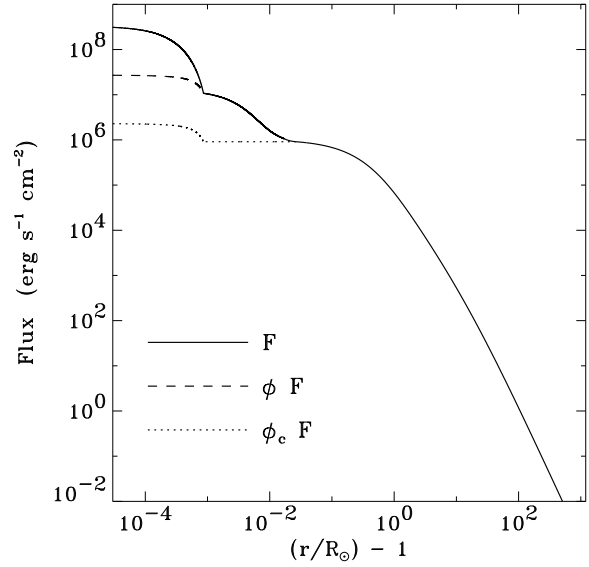}
\caption{\small
Energy flux density of frequency-integrated waves for the
$\sigma_{j} = 3$ km~s$^{-1}$ model: flux within the MBP tubes
({\em{solid line}}), averaged over granular spatial scales
({\em{dashed line}}), and averaged over the supergranule
funnel/canopy structure ({\em{dotted line}}).
The slope discontinuities at $z_{m}$ (i.e.,
$r/R_{\odot} - 1 = 8.6 \times 10^{-4}$) are due to the
discontinuity in $d B_{0} / dr$ at the merging height.}
\end{figure}

\clearpage

\begin{figure}
\epsscale{0.5}
\plotone{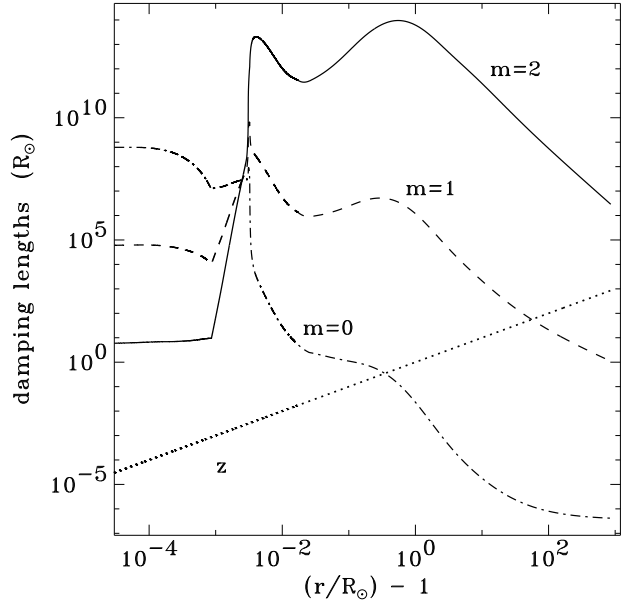}
\caption{\small
Viscous damping length scales plotted versus height for three
assumptions about the effective viscous timescale:
classical $m=0$ viscosity ({\em{dot-dashed line}}),
Williams' (1995) $m=1$ viscosity ({\em{dashed line}}), and
our transverse diffusion $m=2$ viscosity ({\em{solid line}}).
Also plotted is the local height $z$ ({\em{dotted line}}).}
\end{figure}

\begin{figure}
\epsscale{0.5}
\plotone{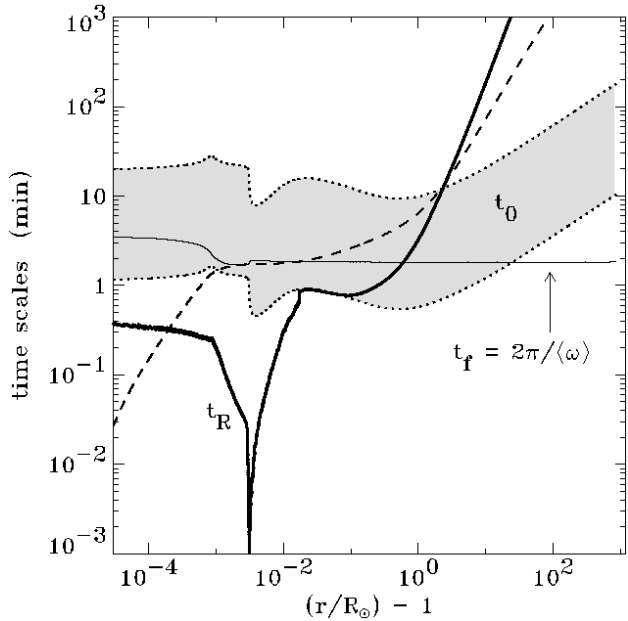}
\caption{\small
Comparison of approximate timescales important for nonlinear
damping: $t_R$ is the wave reflection time ({\em{thick solid
line}}), $t_f$ is the spectrum-weighted wave period ({\em{thin solid
line}}), $t_{\rm trav}$ is the total wave travel time from the
photosphere ({\em{dashed line}}), and $t_0$ ({\em{gray region}})
is a nonlinear turbulent driving time computed for
$\Lambda = 0.06$ ({\em{bottom dotted line}}) and
$\Lambda = 1$ ({\em{top dotted line}}).}
\end{figure}

\clearpage

\begin{figure}
\epsscale{1.00}
\plotone{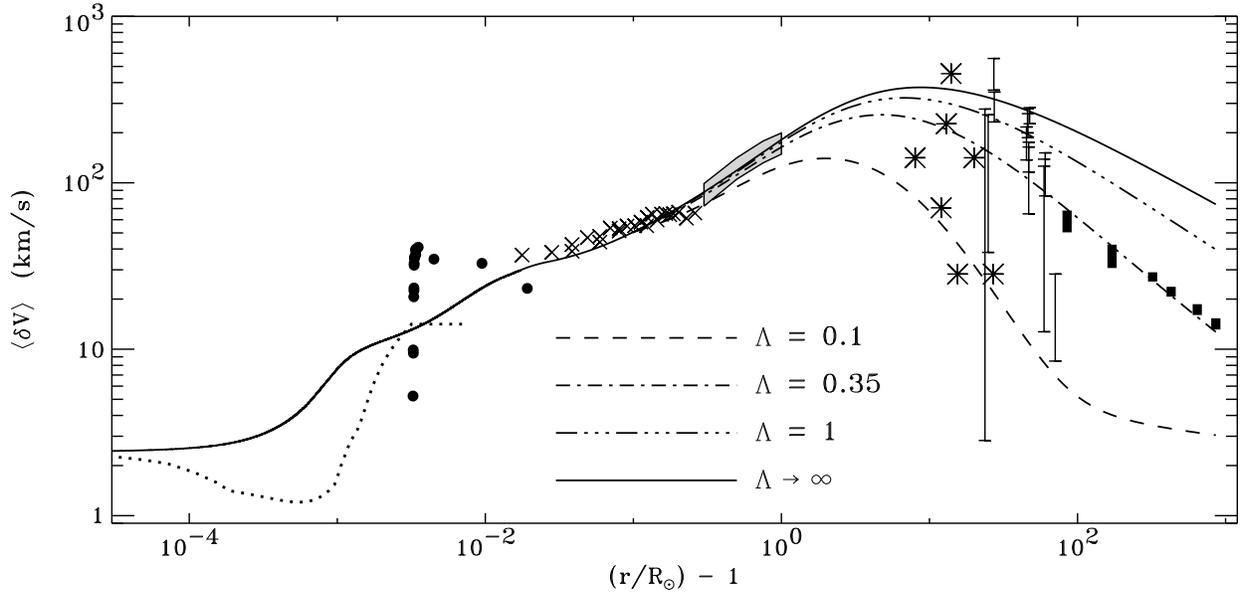}
\caption{\small
Height dependence of the frequency-integrated velocity
amplitude for $\sigma_{j} = 3$ km s$^{-1}$ and a range of
values of the dimensionless outer-scale length-scale
normalization constant $\Lambda$:
0.1 ({\em{dashed line}}),
0.35 ({\em{dash-dotted line}}),
1 ({\em{dash-triple-dotted line}}),
and no damping ({\em{solid line}}).
Other lines and symbols correspond to observations
discussed in detail in {\S}~6.1.}
\end{figure}

\begin{figure}
\epsscale{0.5}
\plotone{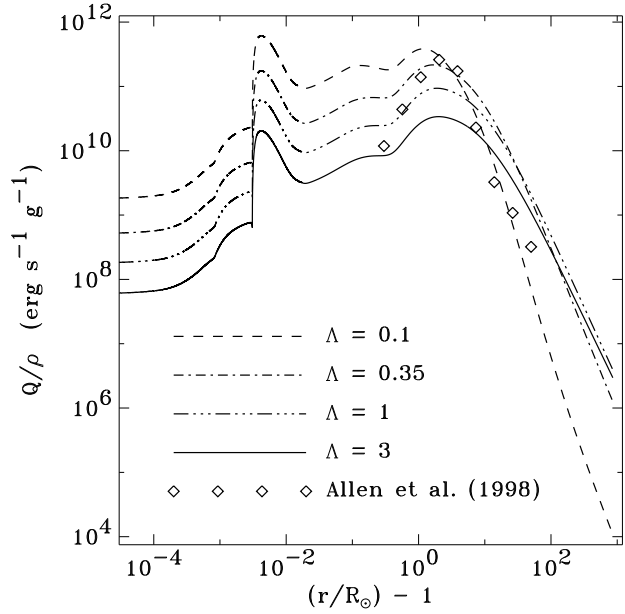}
\caption{\small
Turbulent heating rate per unit mass for the models shown
in Figure 15.  The same line-styles are used for
$\Lambda = 0.1$, 0.35, and 1, and we also show the weaker
damping case $\Lambda = 3$ ({\em{solid line}}).
The diamonds illustrate the empirically constrained heating
rate used by Allen et al.\  (1998).}
\end{figure}

\clearpage

\begin{figure}
\epsscale{0.5}
\plotone{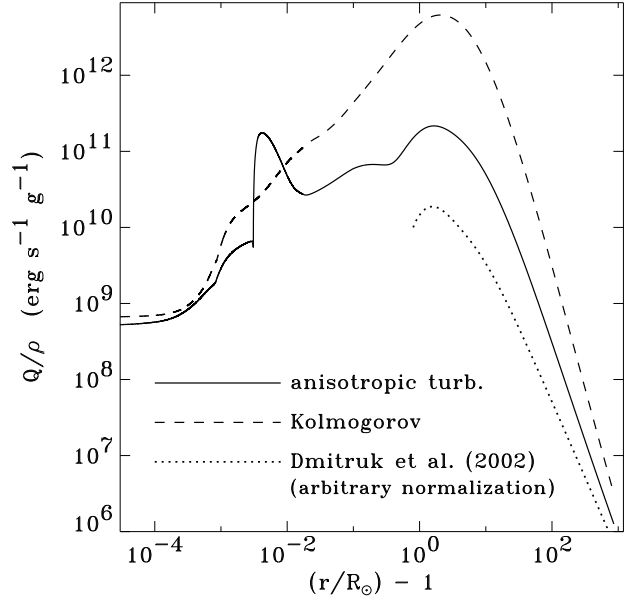}
\caption{\small
Turbulent heating rate per unit mass for the $\Lambda=0.35$
model ({\em{solid line}}) and for the analogous isotropic
Kolmogorov heating rate ({\em{dashed line}}).
The radial dependence of the Dmitruk et al.\  (2002)
analytic formula is also shown with an arbitrary normalization
({\em{dotted line}}).}
\end{figure}

\begin{figure}
\epsscale{0.5}
\plotone{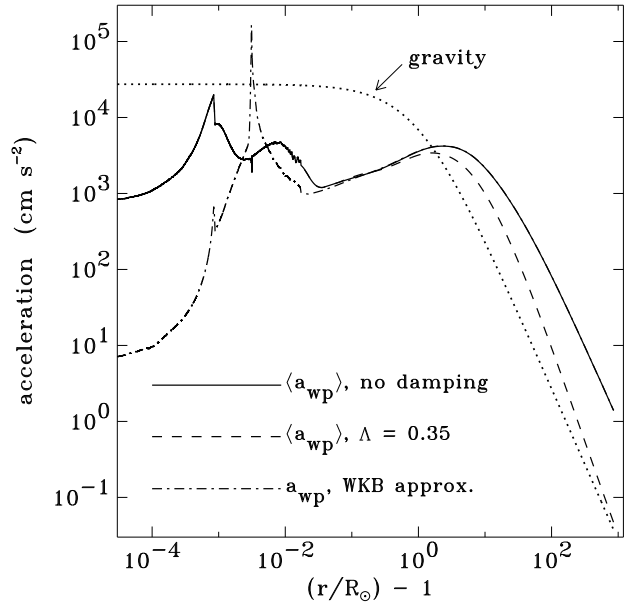}
\caption{\small
Wave pressure acceleration versus height for the undamped
$\sigma_{j} = 3$ km s$^{-1}$ model ({\em{solid line}}),
for the $\Lambda=0.35$ damped model ({\em{dashed line}}),
and for the undamped model computed using the WKB approximation
for the acceleration ({\em{dot-dashed line}}).
The magnitude of the Sun's gravitational acceleration is
also plotted ({\em{dotted line}}).}
\end{figure}

\clearpage

\begin{figure}
\epsscale{0.6}
\plotone{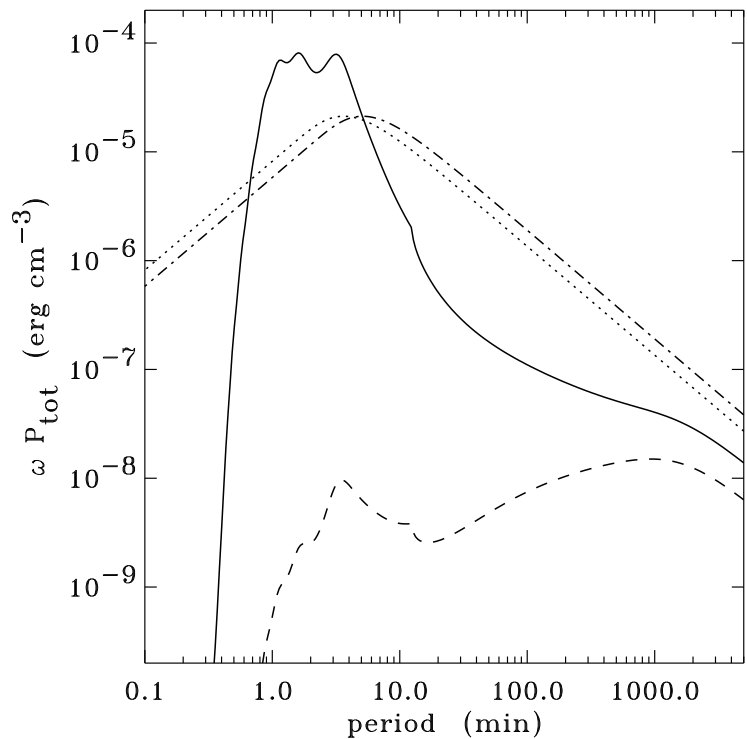}
\caption{\small
Comparison of scaled fluctuation power spectra at 2 $R_{\odot}$.
The linear, empirically constrained frequency spectra
$\omega P_{-}$ ({\em{solid line}}) and $\omega P_{+}$
({\em{dashed line}}) are plotted alongside the nonlinear cascade
frequency spectra projected from 3D wavenumber space at the same
height: $\omega \widetilde{P}_{-}$ ({\em{dotted line}}) and
$\omega \widetilde{P}_{+}$ ({\em{dash-dotted line}}).}
\end{figure}

\end{document}